# Dynamic Exploitation Gaussian Bare-Bones Bat Algorithm for Optimal Reactive Power Dispatch to Improve the Safety and Stability of Power System

Zhaoyang Qu[1,2], Yunchang Dong[1*], Sylvère Mugemanyi[1], Tong Yu[3], Xiaoyong Bo[1], Huashun Li[4], Yang Li[1], François Xavier Rugema[1] and Christophe Bananeza[5]

1 School of Electrical Engineering, Northeast Electric Power University, Jilin, China
2 Jilin Engineering Technology Research Center of Intelligent Electric Power Big Data Processing, Jilin, China
3 Guangxi Power Grid Co., Ltd. Electric Power Research Institute, Nanning, China
4 State Grid Jilin Electric Power Co., Ltd. Jilin Power Supply Company, Jilin, China
5 School of Energy and Electrical Engineering, Hohai University, Nanjing, China

* Correspondence: Yunchang Dong, e-mail: 595245700@qq.com

This work was supported by the Key Projects of National Natural Science Foundation of China under Grant 51437003 and the Jilin science and technology development plan project of China under Grant 20180201092GX.

**ABSTRACT** In this paper, a novel Gaussian bare-bones bat algorithm (GBBBA) and its modified version named as dynamic exploitation Gaussian bare-bones bat algorithm (DeGBBBA) are proposed for solving optimal reactive power dispatch (ORPD) problem. The optimal reactive power dispatch (ORPD) plays a fundamental role in ensuring stable, secure, reliable as well as economical operation of the power system. The ORPD problem is formulated as a complex and nonlinear optimization problem of mixed integers including both discrete and continuous control variables. Bat algorithm (BA) is one of the most popular metaheuristic algorithms which mimics the echolocation of the microbats and which has also outperformed some other metaheuristic algorithms in solving various optimization problems. Nevertheless, the standard BA may fail to balance exploration and exploitation for some optimization problems and hence it may often fall into local optima. The proposed GBBBA employs the Gaussian distribution in updating the bat positions in an effort to mitigate the premature convergence problem associated with the standard BA. The GBBBA takes advantages of Gaussian sampling which begins from exploration and continues to exploitation. DeGBBBA is an advanced variant of GBBBA in which a modified Gaussian distribution is introduced so as to allow the dynamic adaptation of exploitation and exploitation in the proposed algorithm. Both GBBBA and DeGBBBA are used to determine the optimal settings of generator bus voltages, tap setting transformers and shunt reactive sources in order to minimize the active power loss, total voltage deviations and voltage stability index. The performance and effectiveness of the proposed algorithms are demonstrated on three power system test bus including the standard IEEE 14-bus, IEEE 57-bus and IEEE 118-bus. The results provided by the proposed algorithms have been compared with the results obtained by other algorithms reported in the literature. Simulation results show that GBBBA and DeGBBBA are robust and effective in solving the ORPD problem.

## I. INTRODUCTION

THE optimal reactive power dispatch (ORPD) is an important tool for a secure, stable, reliable and economical operation of power system [1]-[2]. The ORPD is one of the main sub-problem of the optimal power flow (OPF) calculations [2]-[4] and it has recently gained much attention from the power system researchers [2], [5].

The ORPD problem can be defined as a complex mixed integer nonlinear optimization problem which involves continuous control variables and discrete control variables [6]-[7]. The main goal of the ORPD problem is to determine the optimal settings of control variables in order to minimize the objectives functions including the total active power transmission losses ($P_{LOSS}$), total voltage deviation (TVD) and voltage stability index (VSI) while satisfying a set of equality and inequality constraints [5], [6], [8].

In the past decades, various classical optimization techniques such as linear programming [9]-[10], nonlinear programming [11], dynamic programming [12], quadratic programming [13]-[16], mixed integer programming [17], interior point method [18]-[19], etc have been applied to solve ORPD problem. Reference [20] proposes an effective multi-objective optimal reactive power dispatch approach for power systems by combining multi-objective evolutionary algorithm and integrated decision making. These classical optimization techniques exhibit significant limitations including algorithmic complexity, insecure convergence, local optimality, sensitivity to initial search point, excessive numerical iterations [2], [21]. These techniques are not suitable for solving the ORPD problems due to their incapability in dealing with nonlinear and non-convex constraints and discontinuous functions as well as problems with multiple local minimum points [3], [21], [22].

Metaheuristics are broadly classified into four classes such as evolutionary algorithms (EAs), physics-based algorithm, human-based algorithms and swarm intelligence algorithms (SIs) [23]-[24].

Evolutionary algorithms are inspired by the biological evolutionary behaviors including recombination, mutation, and selection [24]-[25]. The examples of popular EAs are biogeography-based optimizer (BBO), genetic programming (GP) [26], genetic algorithm (GA) [28], differential evolution (DE) [29], evolutionary programming (EP) [30] and evolution strategy (ES) [31].

Physics-based algorithms mimic physical laws in nature. The most popular physics-based algorithms are simulated anneal (SA) [32], gravitational search algorithm (GSA) [33]. Human-based algorithms imitate certain human behaviors [23]. The most popular human-based algorithms are harmony search algorithm (HAS) [34], tabu search (TS) [35]-[36], teaching-learning-based optimization (TLBO) [37], imperialist competitive algorithm (ICA) [38], Seeker optimization algorithm (SOA) [39] and Exchange market algorithm (EMA) [40]. SIs mimic the social behaviors of the animals living in swarms, flocks, schools or herds [23]-[24]. The most popular SIs are particle swarm optimization (PSO) [41], ant colony optimization (ACO) [42], bat algorithm (BA) [43], dragonfly algorithm (DA) [44], grasshopper optimization algorithm (GOA) [45], salp swarm algorithm (SSA) [46], etc.

Nowadays, numerous metaheuristics such as seeker optimization algorithm (SOA) [47], gravitational search algorithm (GSA) [6], [48], [49], moth-flame algorithm (MFA) [50], imperialist competitive algorithm (ICA) [22], [51], gray wolf optimizer (GWO) [52], invasive weed optimization (IWO) [53], harmony search algorithm (HSA) [54], teaching-learning-based optimization (TLBO) [55], bat algorithm (BA) [7], biogeography-based optimization (BBO) [56], particle swarm optimization (PSO) [57]-[58], cuckoo search algorithm (CSA) [59], genetic algorithm (GA) [60]-[62], krill herd algorithm (KHI) [63], differential evolution (DE) [64], bacteria foraging algorithm (BFA) [65], ant-lion optimizer (ALO) [66], exchange market algorithm (EMA) [5], water cycle algorithm (WCA) [67], etc, have been efficiently utilized to solve the ORPD problem.

BA is one of the most popular swarm intelligence algorithms proposed by Xin-She Yang and which is based on simulating the echolocation of micro-bats [43], [68]. Since its introduction in 2010, BA has attracted many researchers thanks to its simplicity, flexibility and searching capability [69]. However, the standard BA faces a problem of premature convergence while tackling continuous complex problems having high dimensions because of the imbalance between exploration and exploitation capabilities [70]. A broad variety of BA variants have been developed in order to address this problem and thus enhance the performance of the standard BA. These BA variants along with the standard BA have been applied in various research fields [7], [71]-[88].

In this paper, the concept of Gaussian bare-bones which employs Gaussian distribution is incorporated in the standard bat algorithm so as to generate new bat positions and thus make a balance between exploration and exploitation capabilities. The proposed GBBBA and its modified version named as dynamic exploitation Gaussian bare-bones bat algorithm (DeGBBBA) are proposed for solving optimal reactive power dispatch (ORPD) problem. Previously, the concept of Gaussian bare-bones has been applied to various metaheuristics such as PSO [89]-[91], DE [92], water cycle algorithm (WCA) [67], TLBO [55], imperialist competitive algorithm (ICA) [51], artificial bee colony (ABC) [93], BFO [94], fruit fly algorithm [95], CS [96], FA [97], etc. In all these works, the utilization of the Gaussian bare-bones concept has ameliorated the performance of the corresponding metaheuristics. The major contribution of this current work is threefold:

- To introduce for the first time the Gaussian bare-bone concept into the standard bat algorithm by employing the Gaussian distribution to allow the proposed GBBBA algorithm to focus on the exploration phase during the starting iterations and focus on the exploitation phase during the last iterations and thus avoiding the premature convergence of the standard BA.
- To further enhance the performance of the GBBBA by utilizing a modified Gaussian distribution that facilitates the dynamic adaptation of exploitation and exploitation in the proposed DeGBBBA algorithm.
- To apply the proposed GBBBA and DeGBBBA for solving the ORPD problem by minimizing the active power loss, TVD and VSI.

The simulation results revealed that the proposed algorithms outperform other compared algorithms in tackling the ORPD problem.

The rest of the paper is organized as follows: The mathematical formulation of the ORPD problem is discussed in Section II. In Section III, the proposed GBBBA and DeGBBBA are described. The implementation of the proposed GBBBA and DeGBBBA on the ORPD problem is described in Section IV. Simulation results are provided and discussed in Section V. Lastly, the present paper is concluded in Section VI.

## II. PROBLEM FORMULATION

The three different objective functions considered in this paper are described as follows:

### A. Objective Functions

The goal of the ORPD problem is to minimize either the active power loss ($P_{LOSS}$) or total voltage deviation (TVD) or voltage stability index (VSI) at the same time satisfying different equality and inequality constraints.

The ORPD problem is mathematically formulated as follows [2], [3], [5]:

$$\text{Minimize } J(x,u) \quad (1)$$

Subject to

$$g(x,u) = 0 \quad (2)$$
$$h(x,u) \leq 0 \quad (3)$$

where $J(x,u)$ signifies the objective function to be minimized, $g(x,u) = 0$ represents the equality constraints, $h(x,u) \leq 0$ represents the inequality constraints.

$x$ is the vector of dependent variables (state vector) which comprises:
- Load bus voltage $V_L$.
- Generator reactive power output $Q_G$.
- Transmission line loading $S_l$.

Thus, the vector $x$ can be expressed as:

$$x^T = [V_{L1}...V_{LNPQ}, Q_{G1}...Q_{GNG}, S_{l1}...S_{lNTL}] \quad (4)$$

where $NG$ is the number of generators; $NPQ$ is the number of $PQ$ buses and $NTL$ is the number of transmission lines. $u$ is the vector of independent variables (control variables) which comprises:
- Generator bus voltages $V_G$ (continuous control variable).
- Transformer tap settings $T$ (discrete control variable).
- Shunt VAR compensation $Q_C$ (discrete control variable).

Therefore, $u$ can be formed as follows:

$$u^T = [V_{G1},...,V_{GNG}, Q_{C1}...Q_{CNC}, T_1...T_{NT}] \quad (5)$$

where $NT$ and $NC$ are the number of tap regulating transformers and number of shunt VAR compensators, respectively.

*1) Objective Functions*

The minimization of the active power loss ($P_{LOSS}$) is mathematically expressed as follows [2]:

$$\text{Minimize } J_1(x_1,u_1) = \text{minimize } P_{Loss}$$
$$= \sum_{k=0}^{NTL} g_k \left( V_i^2 + V_j^2 - 2V_iV_j \cos\delta_{ij} \right) \quad (6)$$

Where $J_1(x_1,u_1)$ represents the active power loss minimization function of the transmission network, $g_k$ is the conductance of the branch $k$, $V_i$ and $V_j$ are the voltages of $i$ th and $j$ th bus, respectively, $NTL$ is the number of transmission lines, $\delta_{ij}$ is the phase difference of voltages between bus $i$ and $j$.

*2) Minimization of Total Voltage Deviations*

The aim of minimizing the voltage deviations at load buses is to improve the voltage profile along with the security of the electric power network. The mathematical formulation of the total voltage deviations (TVD) at load buses is expressed as [5]:

$$\text{Minimize } J_2(x_2,u_2) = \text{minimize } TVD$$
$$= \sum_{i=1}^{NPQ} \left| V_i - V_i^{ref} \right| \quad (7)$$

where $J_2(x_2,u_2)$ is the total voltage deviation minimization function, $i$ is the element of $NPQ$, $V_i^{ref}$ is the reference voltage magnitude at $i$ th load bus which is taken as 1 $p.u.$

*3) Improvement of Voltage Stability Index*

The voltage stability of the power system is expressed in terms of an index which is referred to as $L-$index/stability index. Its values vary from 0 to 1. The values of the $L-$index closer to 0 denote that the voltage stability of the power system is more stable whereas the values of the $L-$index closer to 1 denote that the voltage stability of the power system is more unstable [5].

The voltage stability index (VSI) or $L_{max}$ can be mathematically expressed as [5]:

$$\text{Minimize } J_3(x_3,u_3) = \text{minimize } L_{max} = \min[\max(L_j)],$$
$$j = 1,2,3,...,NPQ \quad (8)$$

where $L_j$ denotes the voltage stability indicator ($L-$index) of $j$ th node.

The value of $L_j$ is formulated as

$$L_j = \left| 1 - \sum_{i=1}^{NPV} F_{ij} \frac{V_i}{V_j} \right| \quad (9)$$

$$F_{ij} = -[Y_1]^{-1}[Y_2] \quad (10)$$

where $i$ and $j$ represent the elements of $PV$ (Generator) and $PQ$ (Load) buses, respectively. $[Y_1]$ and $[Y_2]$ represent the sub-matrices of the system $Y$ bus which result from the separation of $PV$ and $PQ$ buses as expressed in the following equation:

$$\begin{bmatrix} I_{PQ} \\ I_{PV} \end{bmatrix} = \begin{bmatrix} Y_1 & Y_2 \\ Y_3 & Y_4 \end{bmatrix} \begin{bmatrix} V_{PQ} \\ V_{PV} \end{bmatrix} \quad (11)$$

**B. Constraints**

*1) Equality Constraints*

In (2), $g$ represents the set of equality constraints that describes the load flow equations as follows:

$$P_{Gi} - P_{Di} - V_i \sum_{j=1}^{NB} V_j (G_{ij} \cos \delta_{ij} + B_{ij} \sin \delta_{ij}) = 0 \quad (12)$$

$$Q_{Gi} - Q_{Di} - V_i \sum_{j=1}^{NB} V_j (G_{ij} \cos \delta_{ij} + B_{ij} \sin \delta_{ij}) = 0 \quad (13)$$

where $NB$ signifies the number of buses, $P_{Gi}$ and $Q_{Gi}$ are active and reactive power generation at the $i$ th bus, $P_{Di}$ and $Q_{Di}$ are active and reactive power demands at the $i$ th bus, $G_{ij}$ and $B_{ij}$ are the transfer conductance and susceptance between the $i$ th and the $j$ th buses respectively.

*2) Inequality Constraints*

In (3), $h$ represents the set of inequality constraints comprising:

    *i.    Generator Constraints*

The generation bus voltages including slack bus and reactive power outputs comprising slack bus must be restrained within their lower and upper limits as:

$$V_{Gi}^{\min} \leq V_{Gi} \leq V_{Gi}^{\max}, \quad i = 1, 2, \ldots, NG \quad (14)$$

$$Q_{Gi}^{\min} \leq Q_{Gi} \leq Q_{Gi}^{\max}, \quad i = 1, 2, \ldots, NG \quad (15)$$

where $V_{Gi}^{\min}$ and $V_{Gi}^{\max}$ represent the minimum and maximum generator voltage of $i$ th generating unit, $Q_{Gi}^{\min}$ and $Q_{Gi}^{\max}$ represent the minimum and maximum reactive power output of $i$ th generating unit.

    *ii.    Transformer Constraints*

Transformer tap settings are constrained within their lower and upper limits as follows:

$$T_i^{\min} \leq T_i \leq T_i^{\max}, \quad i = 1, 2, \ldots, NT \quad (16)$$

where $T_i^{\min}$ and $T_i^{\max}$ denote the minimum and maximum tap setting limits of $i$ th transformer.

    *iii.    Shunt VAR Compensator Constraints*

The lower and upper limits of the shunt VAR compensators are obtained by:

$$Q_{Ci}^{\min} \leq Q_{Ci} \leq Q_{Ci}^{\max}, \quad i = 1, 2, \ldots, NC \quad (17)$$

where $Q_{Ci}^{\min}$ and $Q_{Ci}^{\max}$ indicate the minimum and maximum VAR injection limits of $i$ th shunt compensator.

    *iv.    Security Constraints*

The security constraints comprise the constraints on voltages at load buses and transmission line loadings as expressed below.

$$V_{Li}^{\min} \leq V_{Li} \leq V_{Li}^{\max}, \quad i = 1, 2, \ldots, NPQ \quad (18)$$

$$S_{li} \leq S_{li}^{\max}, \quad i = 1, 2, \ldots, NTL \quad (19)$$

where $V_{Li}^{\min}$ and $V_{Li}^{\max}$ represent the minimum and maximum load voltages of $i$ th unit. $S_{li}$ indicates apparent power flow of $i$ th branch and $S_{li}^{\max}$ indicates maximum apparent power flow limit of $i$ th branch.

*3) Constraints Handling*

In handling the ORPD problem, the independent variables are self-constrained in contrast to the dependent variables which are constrained by utilizing the penalty functions. To this end, the equation (1) becomes [5]:

Minimize

$$F = F_{obj} + \lambda_V \sum_{i=1}^{NPQ} (V_{Li} - V_{Li}^{\lim})^2 + \lambda_Q \sum_{i=1}^{NG} (Q_{Gi} - Q_{Gi}^{\lim})^2$$

$$+ \lambda_S \sum_{i=1}^{NTL} (S_{li} - S_{li}^{\lim})^2 \quad (20)$$

where $F_{obj}$ signifies the objective function under consideration i.e. $J_1(x_1, u_1)$ or $J_2(x_2, u_2)$ or $J_3(x_3, u_3)$. $\lambda_V$, $\lambda_G$ and $\lambda_S$ represent the penalty factors related to load bus voltage limit violation, generation reactive power limit violation and line flow violation, respectively. $V_{Li}^{\lim}$, $Q_{Gi}^{\lim}$ and $S_{li}^{\lim}$ are described as the limit values related to load bus voltage, real generation reactive power, and line flow, respectively. They are defined as follows

$$V_{Li}^{\lim} = \begin{cases} V_{Li}^{\max} & if \quad V_{Li} > V_{Li}^{\max} \\ V_{Li}^{\min} & if \quad V_{Li} < V_{Li}^{\min} \\ V_{Li} & if \quad V_{Li}^{\min} \leq V_{Li} \leq V_{Li}^{\max} \end{cases} \quad (21)$$

$$Q_{Gi}^{\lim} = \begin{cases} Q_{Gi}^{\max} & if \quad Q_{Gi} > Q_{Gi}^{\max} \\ Q_{Gi}^{\min} & if \quad Q_{Gi} < Q_{Gi}^{\min} \\ Q_{Gi} & if \quad Q_{Gi}^{\min} \leq Q_{Gi} \leq Q_{Gi}^{\max} \end{cases} \quad (22)$$

$$S_{li}^{\lim} = \begin{cases} S_{li}^{\max} & if & S_{li} > S_{li}^{\max} \\ S_{li}^{\min} & if & S_{li} < S_{li}^{\min} \\ S_{li} & if & S_{li}^{\min} \leq S_{li} \leq S_{li}^{\max} \end{cases} \quad (23)$$

## III. GAUSSIAN BARE-BONES BAT ALGORITHM AND DYNAMIC EXPLORATION GAUSSIAN BARE-BONES BAT ALGORITHM

### A. Brief Review of the Basic Bat Algorithm

The BA simulates the echolocation behavior of the microbats while hunting their prey [68].
The idealization of microbats characteristics has been suggested through the following three rules [68]:
1) All bats employ echolocation to sense the distance between prey and surroundings.
2) All bats fly randomly with velocity $V^i$ at position $X^i$ with a fixed frequency $f^{\min}$, varying wavelength $\lambda$ and loudness $A^0$ to find their prey. The wavelength (or frequency) of the pulses emitted by the bats and the rate of pulse emission $r_1 \in [0,1]$ are tuned in an automatic way according to the proximity of their target.
3) It is supposed that the loudness decreases from a large (positive) $A^0$ to a minimum constant value $A^{\min}$.

Every bat $i$ has a position $X^i$, a velocity $V^i$, a frequency $f^i$ in a $d-$dimensional space. The updated position $X^i$, velocity $V^i$ and frequency $f^i$ are given by the following equations:

$$f^i = f^{\min} + r_1(f^{\max} - f^{\min}) \quad (24)$$
$$V^i(t+1) = V^i(t) + f^i(X^i(t) - X^{best}(t)) \quad (25)$$
$$X^i(t+1) = X^i(t) + V^i(t+1) \quad (26)$$

where $r_1$ is a uniformly distributed random number in the range $[0,1]$; $f^{\min}$ and $f^{\max}$ denote the minimum and maximum allowable frequencies while $f^i$ represents the frequency for the $i$ th bat. In this work, the values of $f^{\min}$ and $f^{\max}$ are set to 0 and 100, respectively as given in [68] and in [77], $t$ is the current iteration number, $X^{best}$ is the location (solution), that has the best fitness in the current population. At initialization, $V^i$ is supposed to be 0.
A new solution for each bat can be obtained locally by means of a random walk as follows.

$$X^{i,new}(t) = \begin{cases} X^{best}(t) + r_3 A^i(t) & if & r_2 > R^i(t) \\ X^r(t) + r_3 A^i(t) & else \end{cases} \quad (27)$$

where $R^i(t)$ denotes the pulse emission rate, $A^i(t)$ denotes the loudness, $r_2 \in [0,1]$, $r_3 \in [-1,1]$, $r \in [1,2,\ldots,N_b]$, $r \neq i$ is a randomly chosen integer, $N_b$ indicates the number of bats or solutions, $X^r(t)$ is a randomly chosen solution in the current iteration, and different from the $i$ th iteration. The fitter solution is given by:

$$X^i(t) = X^{i,new}(t), \text{ if } f(X^{i,new}(t)) < f(X^i(t)), \text{ and}$$
$$r_4 < A^i(t) \; \forall i, \; i \in [1,2,\ldots,N_b] \quad (28)$$

where $r_4 \in [0,1]$ is a uniformly distributed random number. When the prey is found by the bat, the loudness continuously decreases whereas the pulse rate emission continuously increases. The loudness $A^i$ and the pulse emission rate $R^i$ are updated in an iterative way as follows:

$$A^i(t+1) = \alpha A^i(t) \quad (29)$$
$$R^i(t+1) = R^i(0)[1 - \exp(-\gamma t)] \quad (30)$$

where $A^i(0) \in [1,2]$ and $R^i(0) \in [0,1]$ are randomly generated within their respective limits. For the simplification purposes, we set $\alpha = \gamma = 0.9$, as in [66] and in [90].

### B. Bare-Bones Particle Swarm Algorithm (BBPSO)

The PSO is one of the most popular metaheuristics and mimics the social behavior of the birds flocking and fish schooling [41]. The authors in [98]-[99] analyzed the particle trajectories in PSO and have proved that each particle converges to the weighted average of $P^{best}$ and $G^{best}$ as follows:

$$\lim_{t \to +\infty} X^i(t) = \frac{c_1 P^{best,i}(t) + c_2 G^{best}(t)}{c_1 + c_2} \quad (31)$$

where $P^{best}$ is the personal best position, $G^{best}$ is the global best position, $c_1$ and $c_2$ are two learning factors in PSO. According to the convergence rule of PSO, a bare-bones particle swarm optimization (BBPSO) was proposed by Kennedy in [89]. BBPSO does not employ the velocity component to update the positions of the particles. In BBPSO, the positions of the particles are updated by

$$X^{i,d}(t+1) = \begin{cases} N\left(\frac{G^{best}(t) + P^{best,i}(t)}{2}, \left|G^{best}(t) - P^{best,i}(t)\right|\right) & if \; p > 0.5 \\ P^{best,i}(t) & otherwise \end{cases} \quad (32)$$

where $p \in [0,1]$, $N$ represents a Gaussian distribution, $\left(\frac{G^{best}(t) + P^{best,i}(t)}{2}\right)$ denotes the mean or expectation of the distribution, $\left(\left|G^{best}(t) - P^{best,i}(t)\right|\right)$ denotes the standard deviation of the distribution.

In BBPSO, the Gaussian distribution is an adjusting technique that tries to make a fine trade-off between exploration and exploitation trends and this concept can also be employed to improve other metaheuristics [89]. An improved version of BBPSO with Gaussian or Cauchy jumps was proposed in [90].

## C. Gaussian Bare-Bones Bat Algorithm

Inspired by [89], the Gaussian bare-bones bat algorithm (GBBBA) is proposed in this work. The position of each bat $i$ in GBBBA is updated by:

$$X^{i,d}(t+1)=\begin{cases} N\left(\dfrac{X^{best}(t)+X^{i}(t)}{2}, \left|X^{best}(t)-X^{i}(t)\right|\right) & if \quad p>0.5 \\ X^{i}(t) & otherwise \end{cases} \quad (33)$$

The flowchart of the Gaussian bare-bones bat algorithm (GBBBA) is given in Fig.1.

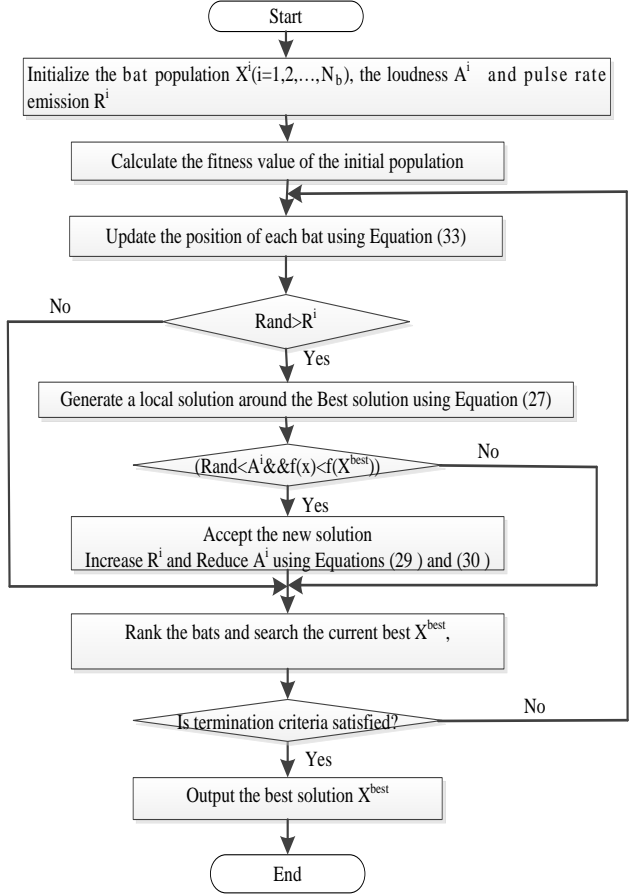

Fig. 1. The Flowchart of GBBBA.

## D. Dynamic Exploitation Gaussian Bare-Bones Bat Algorithm

The authors in [91] proposed a new version of BBPSO referred to as dynamic exploitation BBPSO (DBBePSO) and its hybrid version (HDBBePSO). In DBBePSO and HDBBePSO, a modified Gaussian distribution is employed to dynamically update the exploration and the exploitation throughout the generations. Inspired by this concept in [91], the dynamic exploitation Gaussian bare-bones bat algorithm (DeGBBBA) is proposed in this research. The position of each bat $i$ in DeGBBBA is updated by:

$$X^{i,d}(t+1)=\begin{cases} N\left(\dfrac{X^{best}(t)+X^{i}(t)}{2}, \lambda\left|X^{best}(t)-X^{i}(t)\right|\right) & if \quad p>0.5 \\ X^{i}(t) & otherwise \end{cases} \quad (34)$$

where $\lambda$ is the multiplicative factor in the standard deviation given by:

$$\lambda = 0.9\left(\dfrac{T-t}{T}\right)+0.1 \quad (35)$$

$T$ is the maximum number of iterations and $t$ is the current number of iterations. $\lambda$ exhibits a dynamic variation throughout the iterations commencing from 1 and terminating at 0.1, and thus allowing the exploration at early iterations and then the exploitation at later iterations of the search process.

The flowchart of the dynamic exploitation Gaussian bare-bones bat algorithm (DeGBBBA) is the same as that of GBBBA illustrated in Fig. 1 except that the equation (33) is replaced by the equation (34).

## IV. IMPLEMENTATION OF GBBBA AND DEGBBBA TO ORPD PROBLEM

This section discusses the procedure of applying GBBBA and DeGBBBA to solve ORPD problem.

### A. GBBBA

**Step 1**: The initial position (bat solutions) $X^i$ of every bat is randomly generated between its respective lower and upper limit values. The bat solution $X^i$ represents a vector of the control variables of the ORPD problem including generator voltages, tap setting transformers and shunt compensators.

Initialize the loudness $A^i$ and pulse rate emission $R^i$ and also specify their respective lower and upper limit values.

**Step 2**: Calculate the fitness values of all the bats using the objective function of the problem defined in (6) or (7) or (8) according to the results yielded by New-Raphson power flow analysis [100].

**Step 3**: Update the position of every bat according to (33).

**Step 4**: Generate a new solution by flying randomly according to (27).

**Step 5**: Select the best solution from the old and new solutions, with a probability $A^i(t)$, in accordance with (28).

**Step 6**: Update the values of $A^i$ and $R^i$ according to (29) and (30), respectively.

**Step 7**: Check for the equality constraints of the problem (12)-(13) and the inequality constraints (14)-(19) of the problem.

**Step 8**: Go to step 2 until stopping criterion is satisfied.

### B. DeGBBBA

The application of DeGBBBA to ORPD problem follows almost similar steps as those of GBBBA with the only difference that the position of each bat in Step 3 is updated by the equation (34).

## V. RESULTS AND DISCUSSION

The proposed GBBBA and DeGBBBA have been applied to solve ORPD problems for standard IEEE 14-bus, IEEE 57-bus and IEEE 118-bus test systems. In all three test systems, three different objective functions including the minimization of the active power loss, minimization of the total voltage deviations and enhancement of the voltage stability index have been considered.

The GBBBA and DeGBBBA algorithms have been coded in MATLAB 2014a incorporated with MATPOWER 6.0 [101] and run on PC i.e Lenovo Ideapad 100-15IKB, 2.1 GHz Intel Pentium with 8GB RAM.

The population size is set to 120 ($N_b = 120$). For IEEE 14-bus test system, the number of iterations has been set to 100; for IEEE 14-bus test system, the number of iterations has been set to 100; 200 for IEEE 57-bus test systems and 300 for IEEE 118-bus test systems. In this study, 50 test runs were performed for all the test cases. The results of the DeGBBBA are bold faced for each objective function in order to reveal its optimization capability.

### A. Test System 1: IEEE 14-Bus Test System

The IEEE 14-bus test system is considered as the test system 1. This test system contains five generators at the buses 1, 2, 3, 6, and 8; 20 branches in which 17 branches are transmission lines and 3 branches are tap changing transformers; and 2 shunt VAR compensators installed at buses 9 and 14. In total, this system consists of 10 control variables such as five generators, three tap changing transformers and two shunt capacitors. The detailed data for this test system are given in [55], [64], [102].

The total demands of the system are: $P_{load} = 259 MW$ (active power demand), $Q_{load} = 73.5 MVAR$ (reactive power demand).

The initial total generations and power losses are as follows:
$\sum PG = 272.39 MW$ (active power of generators),
$\sum QG = 82.44 MVAR$ (reactive power of generators),
$P_{LOSS} = 13.49 MW$ (active power losses),
$Q_{LOSS} = -54.54 MVAR$ (reactive power losses).

The control variable limits in p.u. are tabulated in Table 1 [55], [102].

TABLE 1.
CONTROL VARIABLE LIMITS FOR IEEE 14-BUS TEST SYSTEMS [55], [102].

| $V_G^{max}$ | $V_G^{min}$ | $V_{PQ}^{max}$ | $V_{PQ}^{min}$ | $T_k^{max}$ | $T_k^{min}$ | $Q_C^{max}$ | $Q_C^{min}$ |
|---|---|---|---|---|---|---|---|
| 1.1 | 0.95 | 1.05 | 0.95 | 1.1 | 0.9 | 0.3 | 0 |

*1) Case 1: Minimization of Active Power Loss*

The proposed GBBBA and DeGBBBA algorithms are applied on the IEEE 14-bus test system for the minimization of the active power loss defined in (6) considering the penalty terms defined in (20).

The best results obtained by the proposed GBBBA and DeGBBBA; and those obtained by other algorithms reported in the literature such as DE [64], MTLA-DDE [102], MGBTLBO [55], SARGA [61], etc are presented in Table 2.

TABLE 2.
OPTIMAL SETTINGS OF CONTROL VARIABLES FOR IEEE 14-BUS TEST SYSTEMS WITH THE MINIMIZATION OF $P_{Loss}$. OBJECTIVE FUNCTION.

| Variable | Base Case (Initial) | IGSA-CSS [6] | DE [64] | MGB-TLBO [55] | PSO [108] | MTLA-DDE[102] |
|---|---|---|---|---|---|---|
| Generator voltage (p.u) | | | | | | |
| $V_{G1}$ | 1.06 | 1.1 | 1.06 | 1.1 | 1.0917 | 1.07531 |
| $V_{G2}$ | 1.045 | 1.076578 | 1.0449 | 1.0791 | 1.0862 | 1.05734 |
| $V_{G3}$ | 1.01 | 1.046787 | 1.0416 | 1.0485 | 1.0876 | 1.02847 |
| $V_{G6}$ | 1.07 | 1.062305 | 1.1 | 1.0552 | 1.0577 | 0.05057 |
| $V_{G8}$ | 1.09 | 1.097861 | 1.1 | 1.0326 | 1.0278 | 1.03535 |
| Transformer tap ratio (p.u) | | | | | | |
| $T_{4-7}$ | 0.9467 | 1.02 | 1.06 | 1.01 | 1.0039 | 1.08 |
| $T_{4-9}$ | 0.9524 | 0.94 | 1.04 | 1.01 | 1.0324 | 0.91 |
| $T_{5-6}$ | 0.9091 | 1.00 | 1.1 | 1.03 | 1.0083 | 1.01 |
| Capacitor banks (p.u) | | | | | | |
| $Q_{C9}$ | 0.18 | 0.050 | 0.3 | 0.18 | 0.0513 | 0.3 |
| $Q_{C14}$ | 0.18 | - | 0.07 | 0.06 | 0.1920 | 0.08 |
| $P_{LOSS}$(MW) | 13.49 | 12.39706 | 12.3106 | 13.239 | 12.9691 | 12.8978 |
| $TVD$ (p.u) | - | - | - | - | - | - |
| L-index(p.u) | - | - | - | - | - | - |

TABLE 2.
CONTINUED.

| Variable | Base Case(Initial) | JAYA [103] | GSAPSO [48] | CBA-IV [7] | GBBBA | DeGBBBA |
|---|---|---|---|---|---|---|
| Generator voltage (p.u) | | | | | | |
| $V_{G1}$ | 1.06 | 1.10 | 1.1 | 1.0921 | 1.0154 | 1.0945 |
| $V_{G2}$ | 1.045 | 1.086 | 1.076853 | 1.0884 | 1.0262 | 1.0848 |
| $V_{G3}$ | 1.01 | 1.057 | 1.046118 | 1.0558 | 1.0982 | 1.0560 |
| $V_{G6}$ | 1.07 | 1.10 | 1.025647 | 1.0325 | 0.9797 | 1.0501 |
| $V_{G8}$ | 1.09 | 1.092 | 1.096356 | 1.0951 | 1.0207 | 1.1000 |

| | Transformer tap ratio (p.u) | | | | | |
|---|---|---|---|---|---|---|
| $T_{4-7}$ | 0.9467 | 1.045 | 0.96 | 1.0055 | 0.9266 | 0.9802 |
| $T_{4-9}$ | 0.9524 | 0.90 | 1.1 | 1.0059 | 0.9679 | 1.0115 |
| $T_{5-6}$ | 0.9091 | 1.003 | 1.04 | 1.0679 | 0.9342 | 1.0841 |
| | Capacitor banks (p.u) | | | | | |
| $Q_{C9}$ | 0.18 | 0.18 | 0.045 | 0.2208 | 0.1749 | 0.0736 |
| $Q_{C14}$ | 0.18 | 0.18 | - | 0.0786 | 0.0536 | 0.0386 |
| $P_{LOSS}$(MW) | 13.49 | 12.3192 | 12.44901 | 12.2923 | 12.2892 | **12.2864** |
| TVD (p.u) | - | - | - | 0.5308 | 0.3377 | 0.5490 |
| L-index(p.u) | - | - | - | 0.1064 | 0.0420 | 0.1235 |

From the simulation results in Table 2, it is observed that the best (minimum) power loss $P_{Loss}$ achieved by the proposed DeGBBBA algorithm is **12.2864 MW** which also outperforms the other algorithms compared with it.

The comparison of statistical results including the best active power loss (Best), the worst active power loss (Worst), the mean power loss (Mean), the standard deviation (Std) and the percent of power loss reduction (% P save); obtained by different algorithms available in the literature for IEEE 14- bus system are exhibited in Table 3.

TABLE 3.
STATISTICAL RESULTS FOR IEEE 14-BUS TEST SYSTEM WITH THE MINIMIZATION OF $P_{LOSS}$. OBJECTIVE FUNCTION.

| Algorithm | Best (MW) | Worst (MW) | Mean (MW) | Std | % Psave |
|---|---|---|---|---|---|
| PSO-CM [109] | 13.2634 | 13.3142 | 13.2671 | 0.00009 | 1.68 |
| DE [64] | 13.239 | 13.275 | 13.25 | 0.000161 | 1.86 |
| PSO-AM [109] | 13.2371 | 13.2550 | 13.2395 | 0.00006 | 1.87 |
| SARGA [61] | 13.21643 | 13.23891 | 13.22317 | 0.000024 | 2.03 |
| PSO [108] | 12.9691 | - | - | - | 3.86 |
| MTLA-DDE [102] | 12.8978 | 12.8986 | 12.8982 | 0.000006486 | 4.39 |
| GSAPSO [48] | 12.44901 | - | - | - | 7.05 |
| DEEP [107] | 12.4489 | 12.4507 | 12.4494 | 0.0005 | 7.717 |
| DE [106] | 12.4486 | 12.4496 | 12.4486 | 0.0018 | 7.72 |
| CSSP4 [105] | 12.4087 | 12.4974 | 12.4393 | 0.228 | 8 |
| IGSA-CSS [6] | 12.39706 | 12.90281 | 12.46443 | 0.094 | 8.1 |
| DE-ABC [104] | 12.3712 | 12.3712 | 12.3712 | 7.29E-08 | 8.3 |
| JAYA [103] | 12.3192 | - | - | - | 8.68 |
| BA [7] | 12.3171 | 12.3456 | 12.3230 | 0.0099 | 8.7 |
| MGB-TLBO [55] | 12.3106 | 12.3111 | 12.3106 | 0.0000086 | 8.74 |
| CBA-III [7] | 12.3092 | 12.3212 | 12.3161 | 0.0034 | 8.75 |
| CBA-IV [7] | 12.2923 | 12.3098 | 12.3042 | 0.0046 | 8.88 |
| GBBBA | 12.2892 | 12.3036 | 12.30249 | 0.002025 | 8.901 |
| DeGBBBA | **12.2864** | 12.3009 | 12.29318 | 0.002362 | **8.922** |

It is observed from Table 3 that the proposed DeGBBBA algorithm is better than the other algorithms compared with it, because it can yield a power loss reduction of **8.922**% (from the initial power loss) compared with compared with 8.901% by GBBBA, 8.88% by CBA-IV [7], 8.75% by CBA-III [7], 8.74% by MGBTLBO [55], 8.7% by BA [7], 8.68% by JAYA [103], 8.3% by DE-ABC [104], 8.1% by IGSA-CSS [6], 8% by CSSP4 [105], 7.72% by DE [106], 7.717% by DEEP [107], 7.05% by GSAPSO [48], 4.39% by MTLA-DDE [102], 3.86% by PSO [108], 2.03% by SARGA [61], 1.87% by PSO-AM [109], 1.86 by DE [64] and 1.68% by PSO-CM [109].

The proposed DeGBBBA exhibits a better performance over the proposed GBBBA owing to the multiplicative parameter $\lambda$ blended into the Gaussian bare-bones concept as discussed in (34). Fig.2 depicts the convergence characteristics of active power loss (over 100 iterations) achieved by (the best solutions of) the proposed DeGBBBA and GBBBA algorithms for IEEE 14-bus. From this figure, it is seen that the proposed DeGBBBA algorithm provides convergence characteristics than the proposed GBBBA algorithm.

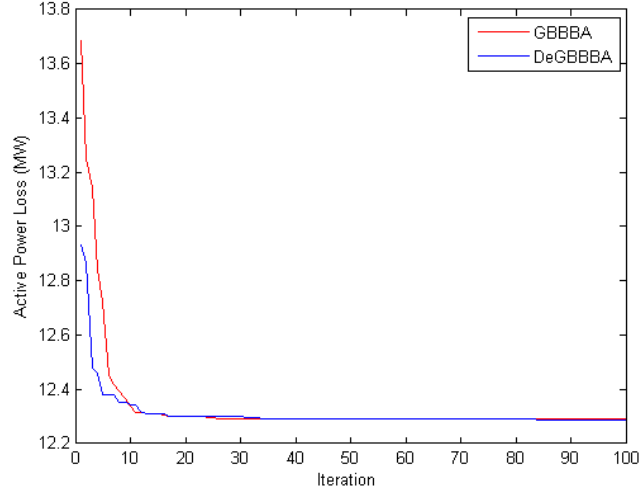

Fig. 2. The convergence characteristics of the GBBBA and DeGBBBA for IEEE 14-bus test power system with the minimization of $P_{Loss}$ objective function.

*2) Case 2: Minimization of Total Voltage Deviation*

The objective function in this case is taken as the minimization of the Total Voltage Deviations (TVD) defined in (7) together with the penalty factors discussed in (20). The best results achieved by the proposed DeGBBBA are shown in Table 4 along with the results achieved by GBBBA, CBA-IV [7], CBA-III [7], BA [7] and IGSA-CSS [6]. It is seen from Table 4 that the proposed DeGBBBA offered a TVD value of **0.032583** in comparison to 0.032621 with GBBBA, 0.0330 with CBA-IV [7], 0.0332 with CBA-III [7], 0.0336 with BA [7] and 0.0339 with IGSA-CSS [6]. According to Table 5, an enhancement of **31.970%** in TVD has been achieved by employing the proposed DeGBBBA compared with the values 31.890%, 31.10%, 30.69%, 29.85% and 29.22% achieved by GBBBA, CBA-IV [7], BA [7], CBA-III [7], BA [7], BA [7], BA [7] and IGSA-CSS [6] respectively. The comparison of the convergence characteristics of TVD for IEEE 14-bus system is illustrated in Fig. 3. From this figure, it is observed that the convergence characteristics of TVD for the proposed DeGBBBA is better than the other algorithms compared with it.

TABLE 4.
OPTIMAL SETTINGS OF CONTROL VARIABLES FOR IEEE 14-BUS TEST SYSTEMS WITH THE MINIMIZATION OF TVD OBJECTIVE FUNCTION.

| Variable | IGSA-CSS [6] | CBA-IV [7] | GBBBA | DeGBBBA |
|---|---|---|---|---|
| Generator voltage (p.u) | | | | |
| $V_{G1}$ | 0.9993 | 0.9958 | 1.0031 | 1.0011 |
| $V_{G2}$ | 1.0004 | 1.0189 | 1.0054 | 1.0018 |
| $V_{G3}$ | 1.0006 | 1.0008 | 0.9967 | 0.9979 |
| $V_{G6}$ | 0.9999 | 1.0102 | 1.0009 | 1.0020 |
| $V_{G8}$ | 0.9993 | 1.0501 | 1.0016 | 0.9996 |
| Transformer tap ratio (p.u) | | | | |
| $T_{4-7}$ | 1.0307 | 1.0121 | 1.0947 | 1.0485 |
| $T_{4-9}$ | 1.0455 | 1.0975 | 1.0281 | 1.0581 |
| $T_{5-6}$ | 0.9966 | 1.0370 | 0.9879 | 0.9268 |
| Capacitor banks (p.u) | | | | |
| $Q_{C9}$ | 0.2308 | 0.0903 | 0.0732 | 0.1259 |
| $Q_{C14}$ | 0.2168 | 0.0637 | 0.2414 | 0.1458 |
| $P_{LOSS}$(MW) | 16.0172 | 16.2499 | 16.2070 | 16.3876 |
| $TVD$ (p.u) | 0.0336 | 0.0330 | 0.032621 | **0.032583** |
| L-index(p.u) | 0.0835 | 0.2514 | 0.1344 | 0.0966 |

TABLE 5.
STATISTICAL RESULTS FOR IEEE 14-BUS TEST SYSTEM WITH THE minimization OF TVD OBJECTIVE FUNCTION.

| Algorithm | Best (p.u.) | Worst (p.u.) | Mean (p.u.) | Std | % TVD Improve |
|---|---|---|---|---|---|
| IGSA-CSS [6] | 0.03390 | 0.09056 | 0.04583 | 0.017 | 29.22 |
| BA [7] | 0.0336 | 0.0515 | 0.0410 | 0.0058 | 29.85 |
| CBA-III [7] | 0.0332 | 0.0489 | 0.0399 | 0.0054 | 30.69 |
| CBA-IV [7] | 0.0330 | 0.0425 | 0.0368 | 0.0029 | 31.10 |
| GBBBA | 0.032621 | 0.037694 | 0.03513686 | 0.000512468 | 31.890 |
| DeGBBBA | **0.032583** | 0.037155 | 0.03512532 | 0.00046485 | **31.970** |

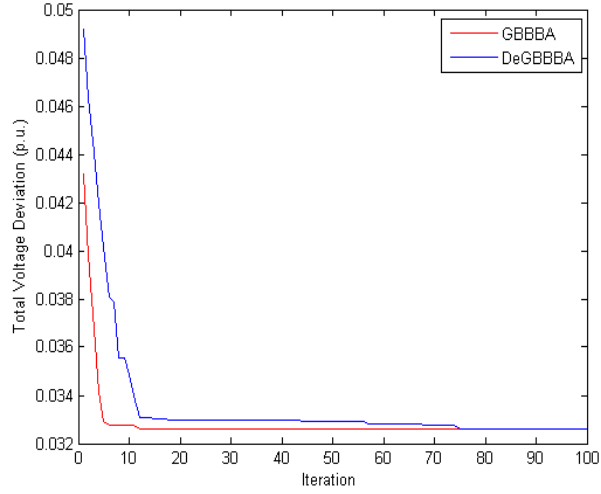

Fig. 3. The convergence characteristics of the GBBBA and DeGBBBA for IEEE 14-bus test power system with the minimization of TVD objective function.

*3) Case 3: Improvement of Voltage Stability Index*

In this case, the proposed algorithms are utilized for minimizing the Voltage Stability Index (VSI) presented in (8) as the objective function together with the penalty terms defined in (20). The best results yielded by the proposed DeGBBBA and GBBBA are reported in Table 6 along with the best results yielded by other algorithms. It is observed from this table that an $L-$ index value of **0.016518** is obtained by the DeGBBBA compared to 0.016507 with the GBBBA, 0.0170 with CBA-IV [7], 0.0174 with CBA-III [7] and 0.0179 with BA [7]. Table 7 summarizes the statistical results for the present case. It is found from Table 7 that the proposed DeGBBA also yields the smallest Best, Mean and Std values compared with other algorithms. Fig. 4 shows the comparison of convergence characteristics of VSI for IEEE 14-bus system. From this figure, it must be acknowledged that the convergence characteristics from the proposed DeGBBBA is better than the convergence characteristics from the proposed GBBBA.

TABLE 6.
OPTIMAL SETTINGS OF CONTROL VARIABLES FOR IEEE 14-BUS TEST SYSTEMS WITH THE MINIMIZATION OF VSI OBJECTIVE FUNCTION.

| Variable | CBA-IV [7] | GBBBA | DeGBBBA |
|---|---|---|---|
| Generator voltage (p.u) | | | |
| $V_{G1}$ | 1.100 | 1.0696 | 1.0154 |
| $V_{G2}$ | 1.0964 | 1.0104 | 1.0262 |
| $V_{G3}$ | 0.9270 | 0.9798 | 1.0982 |
| $V_{G6}$ | 1.0187 | 1.0979 | 0.9797 |
| $V_{G8}$ | 1.0003 | 0.9715 | 1.0207 |
| Transformer tap ratio (p.u) | | | |
| $T_{4-7}$ | 0.9975 | 0.9041 | 0.9266 |
| $T_{4-9}$ | 1.0413 | 1.0885 | 0.9679 |
| $T_{5-6}$ | 0.9495 | 0.9622 | 0.9342 |
| Capacitor banks (p.u) | | | |
| $Q_{C9}$ | 0.0572 | 0.1086 | 0.1749 |
| $Q_{C14}$ | 0.0903 | 0.0246 | 0.0536 |
| $P_{LOSS}$(MW) | 18.3607 | 16.2378 | 15.2731 |
| $TVD$ (p.u) | 0.4459 | 0.3836 | 0.3377 |
| L-index(p.u) | 0.0170 | 0.016507 | **0.016418** |

TABLE 7.
STATISTICAL RESULTS FOR IEEE 14-BUS TEST SYSTEM WITH THE MINIMIZATION OF VSI OBJECTIVE FUNCTION.

| Algorithm | Best (p.u.) | Worst (p.u.) | Mean (p.u.) | Std |
|---|---|---|---|---|
| BA [7] | 0.0179 | 0.0204 | 0.0189 | $7.8736 \times 10^{-4}$ |
| CBA-III [7] | 0.0174 | 0.0187 | 0.0179 | $3.8850 \times 10^{-4}$ |
| CBA-IV [7] | 0.0170 | 0.0182 | 0.0175 | $3.6334 \times 10^{-4}$ |
| GBBBA | 0.016507 | 0.017211 | 0.0170934 | 8.82298E-05 |
| DeGBBBA | **0.016418** | 0.016992 | 0.01689124 | 5.72854E-05 |

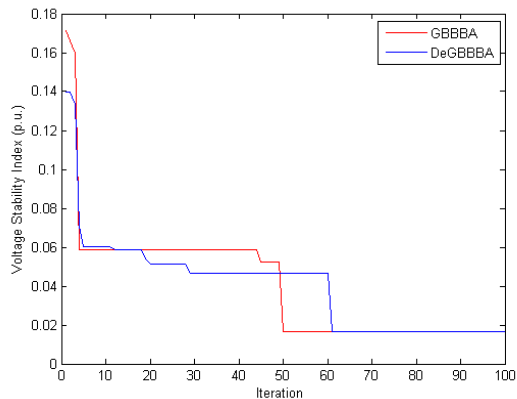

Fig. 4. The convergence characteristics of the GBBBA and DeGBBBA for IEEE 14-bus test power system with the minimization of VSI objective function.

## B. Test System 2: IEEE 57-Bus Test System

The IEEE 57-bus test system is considered as the test system 2. The IEEE 57 bus test system has 7 generators at the buses 1,2,3,6,8,9 and 12; 80 transmission lines; 15 tap changing transformers and 3 shunt VAR compensators at buses 18, 25 and 53. Totally, the IEEE 57-bus test system consists of 25 control variables such as 7 generators, 15 tap changing transformers and 3 shunt capacitors.

See references [47], [63] and [104] for detailed data of this test system. The total demands of the system are:
$P_{load} = 1250.8 MW$ (active power demand),
$Q_{load} = 336.4 MVAR$ (reactive power demand). The initial total generations and power losses are given by:
$\sum PG = 1279.26 MW$ (active power of generators),
$\sum QG = 345.45 MVAR$ (reactive power of generators),
$P_{LOSS} = 28.462 MW$ (active power losses),
$Q_{LOSS} = -124.27 MVAR$ (reactive power losses).
The control variable limits in p.u. are given in Table 8 [104].

TABLE 8.
CONTROL VARIABLE LIMITS FOR IEEE 57-BUS TEST SYSTEMS [104].

| $V_G^{max}$ | $V_G^{min}$ | $V_{PQ}^{max}$ | $V_{PQ}^{min}$ | $T_k^{max}$ | $T_k^{min}$ | $Q_C^{max}$ | $Q_C^{min}$ |
|---|---|---|---|---|---|---|---|
| 1.1 | 0.9 | 1.1 | 0.9 | 1.1 | 0.9 | 0.2 | 0 |

### 1) Case 1: Minimization of Active Power Loss

The best results achieved by different algorithms for minimizing the active power loss of the IEEE 57-bus test system are depicted in Table 9. According to Table 9, the DeGBBBA algorithm achieves **21.9499 MW** active power loss which is better than the active power losses achieved by the other compared algorithms. Table 10 presents the comparative statistical results. It is also found from Table 10 that a **22.880 %** reduction (from the initial loss of 28.462 MW) in active power loss is obtained by employing the DeGBBBA algorithm, which outperforms the values obtained by the other algorithms. Fig. 5 shows the comparative convergence characteristics of active power loss over 200 iterations provided by the proposed algorithms. According to Fig. 6, that the DeGBBBA yields better convergence characteristics than the GBBBA.

TABLE 9.
OPTIMAL SETTINGS OF CONTROL VARIABLES FOR IEEE 57-BUS TEST SYSTEMS WITH THE MINIMIZATION OF $P_{LOSS}$. OBJECTIVE FUNCTION.

| Variable | PSO-ICA [22] | MICA-IWO [53] | CKHA [63] | ABC [110] | NGBWCA [67] | SOA [47] |
|---|---|---|---|---|---|---|
| Generator voltage (p.u) | | | | | | |
| $V_{G1}$ | 1.0395 | 1.06 | 1.0600 | 1.0808 | 1.0600 | 1.06 |
| $V_{G2}$ | 1.0259 | 1.05841 | 1.0590 | 1.0637 | 1.0591 | 1.0580 |
| $V_{G3}$ | 1.077 | 1.04568 | 1.0487 | 1.0467 | 1.0492 | 1.0437 |
| $V_{G6}$ | 0.9982 | 1.03969 | 1.0431 | 1.0337 | 1.0399 | 1.0352 |
| $V_{G8}$ | 1.0158 | 1.06 | 1.0600 | 1.0464 | 1.0586 | 1.0548 |
| $V_{G9}$ | 0.985 | 1.02737 | 1.0447 | 1.0253 | 1.0461 | 1.0369 |
| $V_{G12}$ | 0.9966 | 1.03499 | 1.0410 | 1.0526 | 1.0413 | 1.0336 |
| Transformer tap ratio (p.u) | | | | | | |
| $T_{4-18}$ | 0.9265 | 1.01 | 0.9179 | 0.97 | 0.9712 | 1.00 |
| $T_{4-18}$ | 0.9532 | 0.95 | 1.0256 | 0.94 | 0.9243 | 0.96 |
| $T_{21-20}$ | 1.0165 | 1.02 | 0.9000 | 0.97 | 0.9123 | 1.01 |
| $T_{24-26}$ | 1.0071 | 1.01 | 0.9020 | 0.97 | 0.9001 | 1.01 |
| $T_{7-29}$ | 0.9414 | 0.96 | 0.9104 | 0.94 | 0.9112 | 0.97 |
| $T_{34-32}$ | 0.9555 | 0.98 | 0.9005 | 0.93 | 0.9004 | 0.97 |
| $T_{11-41}$ | 0.9032 | 0.9 | 0.9000 | 0.90 | 0.9128 | 0.90 |
| $T_{15-45}$ | 0.9356 | 0.95 | 0.9000 | 0.99 | 0.9000 | 0.97 |
| $T_{14-46}$ | 0.9172 | 0.94 | 1.0797 | 0.96 | 1.0218 | 0.95 |
| $T_{10-51}$ | 0.9337 | 0.95 | 0.9887 | 0.98 | 0.9902 | 0.96 |
| $T_{13-49}$ | 0.9 | 0.91 | 0.9914 | 0.93 | 0.9568 | 0.92 |
| $T_{11-43}$ | 0.9206 | 0.95 | 0.9000 | 0.92 | 0.9000 | 0.96 |
| $T_{40-56}$ | 1.0042 | 0.1 | 0.9002 | 0.96 | 0.9000 | 1.00 |
| $T_{39-57}$ | 1.0297 | 0.97 | 1.0173 | 0.94 | 1.0118 | 0.96 |
| $T_{9-55}$ | 0.9294 | 0.96 | 1.0023 | 0.94 | 1.0000 | 0.97 |
| Capacitor banks (p.u) | | | | | | |
| $Q_{C18}$ | 0.099846 | 0.1 | 0.0994 | 0.15 | 0.0914 | 0.09984 |
| $Q_{C25}$ | 0.10 | 0.059 | 0.0590 | 0.17 | 0.0587 | 0.05808 |
| $Q_{C53}$ | 0.10 | 0.063 | 0.0630 | 0.13 | 0.0634 | 0.06288 |
| $P_{LOSS}$(MW) | 25.5856 | 24.25684 | 23.38 | 23.9666 | 23.27 | 24.26548 |
| TVD (p.u) | - | - | - | - | - | - |
| L-index(p.u) | - | - | - | - | - | - |

TABLE 9.
CONTINUED.

| Variable | MGBICA [51] | ALC-PSO [111] | BBO [56] | CPVEIHBMO [112] | MFO [50] |
|---|---|---|---|---|---|
| Generator voltage (p.u) | | | | | |
| $V_{G1}$ | 1.06 | 1.0600 | 1.06 | 1.076 | 1.06000 |
| $V_{G2}$ | 1.0492 | 1.0593 | 1.0504 | 1.054 | 1.05870 |
| $V_{G3}$ | 1.0388 | 1.0491 | 1.0440 | 1.035 | 1.04690 |
| $V_{G6}$ | 1.0353 | 1.0432 | 1.0376 | 1.013 | 1.04210 |
| $V_{G8}$ | 1.0558 | 1.0600 | 1.0550 | 1.044 | 1.06000 |

| Variable | | | | | |
|---|---|---|---|---|---|
| $V_{G9}$ | 1.0212 | 1.0451 | 1.0229 | 1.093 | 1.04230 |
| $V_{G12}$ | 1.0295 | 1.0411 | 1.0323 | 0.989 | 1.03730 |
| Transformer tap ratio (p.u) | | | | | |
| $T_{4-18}$ | 0.95 | 0.9611 | 0.96693 | 1.029 | 0.95011 |
| $T_{4-18}$ | 1 | 0.9109 | 0.99022 | 1.034 | 1.00760 |
| $T_{21-20}$ | 1.01 | 0.9000 | 1.0120 | 0.989 | 1.00630 |
| $T_{24-26}$ | 1.02 | 0.9004 | 1.0087 | 1.016 | 1.00760 |
| $T_{7-29}$ | 0.99 | 0.9106 | 0.97074 | 0.994 | 0.97523 |
| $T_{34-32}$ | 0.93 | 0.9000 | 0.96869 | 1.100 | 0.97218 |
| $T_{11-41}$ | 0.91 | 0.9000 | 0.90082 | 1.072 | 0.90000 |
| $T_{15-45}$ | 0.97 | 0.9000 | 0.96602 | 1.000 | 0.97186 |
| $T_{14-46}$ | 0.96 | 1.0275 | 0.95079 | 0.987 | 0.95355 |
| $T_{10-51}$ | 0.96 | 0.9876 | 0.96414 | 0.933 | 0.96736 |
| $T_{13-49}$ | 0.92 | 0.9756 | 0.92462 | 1.029 | 0.92788 |
| $T_{11-43}$ | 0.95 | 0.9000 | 0.95022 | 1.0923 | 0.96406 |
| $T_{40-56}$ | 1.03 | 0.9000 | 0.99666 | 0.996 | 0.99980 |
| $T_{39-57}$ | 0.98 | 1.0121 | 0.96289 | 1.0645 | 0.96060 |
| $T_{9-55}$ | 0.99 | 0.9944 | 0.96001 | 0.9847 | 0.97899 |
| Capacitor banks (p.u) | | | | | |
| $Q_{C18}$ | 0.04 | 0.0994 | 0.09782 | 0.0653 | 0.099968 |
| $Q_{C25}$ | 0.06 | 0.0590 | 0.05904 | 0.0084 | 0.059000 |
| $Q_{C53}$ | 0.03 | 0.0630 | 0.06288 | 0.0763 | 0.063000 |
| $P_{LOSS}$(MW) | 24.8863 | 23.39 | 24.544 | 22.78 | 24.25293 |
| $TVD$ (p.u) | 1.0283 | 1.2697 | - | - | - |
| L-index(p.u) | - | - | - | - | - |

TABLE 9.
CONTINUED.

| Variable | OGSA [8] | GSA [113] | CBA-IV [7] | GBBBA | DeGBBBA |
|---|---|---|---|---|---|
| Generator voltage (p.u) | | | | | |
| $V_{G1}$ | 1.06 | 1.060000 | 1.0964 | 0.9711 | 1.0930 |
| $V_{G2}$ | 1.0594 | 1.060000 | 1.0999 | 1.0561 | 1.0934 |
| $V_{G3}$ | 1.0492 | 1.060000 | 1.0906 | 1.0168 | 1.0944 |
| $V_{G6}$ | 1.0433 | 1.008102 | 1.0838 | 1.0877 | 1.0842 |
| $V_{G8}$ | 1.0600 | 1.054955 | 1.1000 | 1.0156 | 1.0130 |
| $V_{G9}$ | 1.0450 | 1.009801 | 1.0869 | 1.0778 | 1.0874 |
| $V_{G12}$ | 1.0407 | 1.018591 | 1.0822 | 0.9773 | 1.0865 |
| Transformer tap ratio (p.u) | | | | | |
| $T_{4-18}$ | 0.9000 | 1.100000 | 0.9002 | 0.9457 | 0.9229 |
| $T_{4-18}$ | 0.9947 | 1.082634 | 0.9005 | 1.0360 | 1.0651 |
| $T_{21-20}$ | 0.9000 | 0.921987 | 0.9958 | 1.0128 | 1.0180 |
| $T_{24-26}$ | 0.9001 | 1.016731 | 1.0086 | 0.9817 | 1.0452 |
| $T_{7-29}$ | 0.9111 | 0.996262 | 0.9061 | 0.9827 | 1.0027 |
| $T_{34-32}$ | 0.9000 | 1.100000 | 0.9990 | 1.0150 | 0.9795 |
| $T_{11-41}$ | 0.9000 | 1.074625 | 0.9087 | 1.0114 | 0.9119 |
| $T_{15-45}$ | 0.9000 | 0.954340 | 0.9003 | 1.0441 | 0.9623 |
| $T_{14-46}$ | 1.0464 | 0.937722 | 0.9002 | 1.0000 | 0.9680 |
| $T_{10-51}$ | 0.9875 | 1.016790 | 0.9123 | 0.9530 | 0.9912 |
| $T_{13-49}$ | 0.9638 | 1.052572 | 0.9002 | 0.9187 | 0.9552 |
| $T_{11-43}$ | 0.9000 | 1.100000 | 0.9000 | 0.9116 | 1.0441 |
| $T_{40-56}$ | 0.9000 | 0.979992 | 1.0267 | 0.9796 | 1.0181 |
| $T_{39-57}$ | 1.0148 | 1.024653 | 0.9729 | 0.9707 | 1.0046 |
| $T_{9-55}$ | 0.9830 | 1.037316 | 0.9220 | 1.0415 | 1.0262 |
| Capacitor banks (p.u) | | | | | |
| $Q_{C18}$ | 0.0682 | 0.078254 | 0.1827 | 0.0253 | 0.0175 |
| $Q_{C25}$ | 0.0590 | 0.005869 | 0.1335 | 0.0456 | 0.1462 |
| $Q_{C53}$ | 0.0630 | 0.046872 | 0.0858 | 0.1335 | 0.1780 |
| $P_{LOSS}$(MW) | 23.43 | 23.461194 | 21.9627 | 21.9526 | **21.9499** |
| $TVD$ (p.u) | 1.1907 | - | 1.6661 | 1.3285 | 1.6748 |
| L-index(p.u) | 0.4120 | - | 0.1508 | 0.1427 | 0.1396 |

TABLE 10.
STATISTICAL RESULTS FOR IEEE 57-BUS TEST SYSTEM WITH THE MINIMIZATION OF $P_{LOSS}$ OBJECTIVE FUNCTION.

| Algorithm | Best (MW) | Worst (MW) | Mean (MW) | Std | % Psave |
|---|---|---|---|---|---|
| SOA [47] | 24.26548 | 24.28046 | 24.27078 | 4.2081E-005 | 14.7443 |
| MICA-IWO [53] | 24.25684 | 24.28364 | 24.27561 | 2.3361E-004 | 14.7746 |
| ALC-PSO [111] | 23.39 | 23.44 | 23.41 | $82 \times 10^{-5}$ | 17.82 |
| SR-DE [4] | 23.3550 | 24.1656 | 23.4392 | 0.1458 | 17.94 |
| BA [7] | 22.030 | 23.2109 | 22.6017 | 0.3482 | 22.6 |
| CBA-III [7] | 22.0162 | 22.9062 | 22.4721 | 0.2735 | 22.65 |
| CBA-IV [7] | 21.9627 | 22.7961 | 22.0853 | 0.2241 | 22.84 |
| GBBBA | 21.9526 | 22.3604 | 22.0716 | 0.044985 | 22.870 |
| DeGBBBA | **21.9499** | 22.2211 | 22.04412 | 0.040117 | **22.880** |

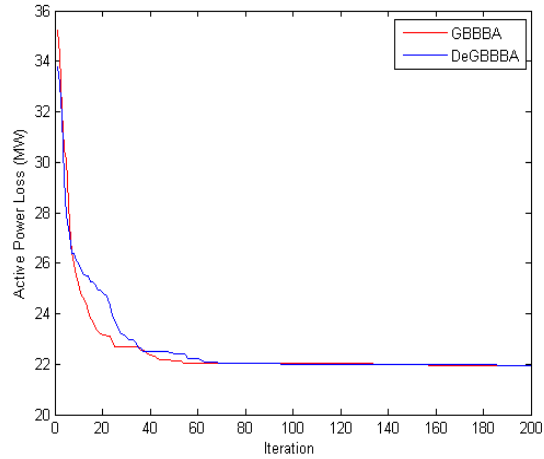

Fig. 5. The convergence characteristics of the GBBBA and DeGBBBA for IEEE 57-bus test power system with the minimization of $P_{LOSS}$ objective function.

*2) Case 2: Minimization of Total Voltage Deviation*

Table 11 gives the best results of TVD minimization for IEEE 57-bus system yielded by the DeGBBBA and GBBA. It is observed from this table that the DeGBBBA is better than other compared algorithms because it provides a TVD value of 0.6311 against 0.6368 with GBBBA, 0.6399 with CBA-IV [7], 0.6413 with CBA-III [7], 0.6434 with BA [7], 0.6501 with NGBWCA [67], 0.6634 with ALC-PSO [111] and 0.6982 with OGSA [8].

TABLE 11.
OPTIMAL SETTINGS OF CONTROL VARIABLES FOR IEEE 57-BUS TEST SYSTEMS WITH THE minimization OF TVD OBJECTIVE FUNCTION.

| Variable | CKHA [63] | MGBICA [51] | OGSA [8] | ALC-PSO [111] |
|---|---|---|---|---|
| Generator voltage (p.u) | | | | |
| $V_{G1}$ | 1..0236 | 1.0555 | 1.0138 | 1.0172 |
| $V_{G2}$ | 1.0121 | 1.0339 | 0.9608 | 0.9819 |
| $V_{G3}$ | 1.0030 | 1.0086 | 1.0173 | 1.0044 |
| $V_{G6}$ | 1.0052 | 1.0067 | 0.9898 | 1.0050 |
| $V_{G8}$ | 1.0180 | 1.0462 | 1.0362 | 1.0208 |
| $V_{G9}$ | 1.0427 | 1.0067 | 1.0241 | 1.0258 |
| $V_{G12}$ | 1.0030 | 1.0059 | 1.0136 | 1.0080 |
| Transformer tap ratio (p.u) | | | | |
| $T_{4-18}$ | 0.9650 | 0.93 | 0.9833 | 1.0197 |
| $T_{4-18}$ | 0.9877 | 1.01 | 0.9503 | 0.9609 |
| $T_{21-20}$ | 0.9584 | 0.98 | 0.9523 | 0.9465 |
| $T_{24-26}$ | 1.0085 | 1.07 | 1.0036 | 0.9923 |
| $T_{7-29}$ | 1.0112 | 0.96 | 0.9778 | 0.9960 |
| $T_{34-32}$ | 0.9000 | 0.91 | 0.9146 | 0.9000 |
| $T_{11-41}$ | 0.9784 | 0.9 | 0.9454 | 0.9622 |
| $T_{15-45}$ | 0.9000 | 0.95 | 0.9265 | 0.9059 |
| $T_{14-46}$ | 0.9809 | 0.95 | 0.9960 | 0.9764 |
| $T_{10-51}$ | 1.0388 | 0.98 | 1.0386 | 1.0622 |
| $T_{13-49}$ | 0.9041 | 0.94 | 0.9060 | 0.9106 |
| $T_{11-43}$ | 0.9119 | 0.97 | 0.9234 | 0.9289 |
| $T_{40-56}$ | 0.9899 | 1.04 | 0.9871 | 0.9771 |
| $T_{39-57}$ | 1.0213 | 0.93 | 1.0132 | 1.0281 |
| $T_{9-55}$ | 0.9078 | 0.98 | 0.9372 | 0.9001 |
| Capacitor banks (p.u) | | | | |
| $Q_{C18}$ | 0.0601 | 0.03 | 0.0463 | 0.0585 |
| $Q_{C25}$ | 0.0590 | 0.06 | 0.0590 | 0.0587 |
| $Q_{C53}$ | 0.0630 | 0.03 | 0.0628 | 0.0381 |
| $P_{LOSS}$(MW) | 28.46 | 26.4618 | 32.34 | 0.2931 |
| TVD (p.u) | 0.6484 | 0.77461 | 0.6982 | 0.6634 |
| L-index(p.u) | 0.1899 | - | 0.5123 | - |

TABLE 11.
CONTINUED.

| Variable | NGBWCA [67] | CBA-IV [7] | GBBBA | DeGBBBA |
|---|---|---|---|---|
| Generator voltage (p.u) | | | | |
| $V_{G1}$ | 1.0151 | 1.0014 | 0.9966 | 0.9127 |
| $V_{G2}$ | 0.9810 | 1.0010 | 0.9983 | 1.0040 |
| $V_{G3}$ | 1.0002 | 1.0002 | 0.9948 | 0.9394 |
| $V_{G6}$ | 1.0039 | 0.9995 | 0.9998 | 1.0980 |
| $V_{G8}$ | 1.0198 | 0.9998 | 1.0026 | 0.9110 |
| $V_{G9}$ | 1.0254 | 1.0003 | 0.9991 | 1.0690 |
| $V_{G12}$ | 1.0081 | 0.9994 | 1.0001 | 0.9416 |
| Transformer tap ratio (p.u) | | | | |
| $T_{4-18}$ | 1.0185 | 0.9638 | 1.0499 | 0.9890 |
| $T_{4-18}$ | 0.9601 | 0.9172 | 1.0318 | 1.0457 |
| $T_{21-20}$ | 0.9458 | 0.9997 | 1.0741 | 1.0684 |
| $T_{24-26}$ | 0.9919 | 0.9418 | 1.0517 | 0.9200 |
| $T_{7-29}$ | 0.9951 | 1.0699 | 0.9151 | 0.9425 |
| $T_{34-32}$ | 0.9000 | 0.9352 | 0.9988 | 1.0523 |
| $T_{11-41}$ | 0.9622 | 1.0632 | 0.9345 | 0.9576 |
| $T_{15-45}$ | 0.9058 | 1.0644 | 0.9821 | 1.0886 |
| $T_{14-46}$ | 0.9764 | 0.9109 | 1.0312 | 1.0196 |
| $T_{10-51}$ | 1.0600 | 0.9224 | 1.0418 | 0.9210 |
| $T_{13-49}$ | 0.9100 | 1.0108 | 1.0375 | 0.9074 |
| $T_{11-43}$ | 0.9302 | 0.9982 | 1.0949 | 0.9269 |
| $T_{40-56}$ | 0.9770 | 1.0809 | 0.9591 | 1.0245 |
| $T_{39-57}$ | 1.0271 | 1.0111 | 0.9022 | 1.0665 |
| $T_{9-55}$ | 0.9000 | 0.9112 | 0.9420 | 1.0528 |
| Capacitor banks (p.u) | | | | |
| $Q_{C18}$ | 0.0550 | 0.0992 | 0.0203 | 0.0073 |
| $Q_{C25}$ | 0.0590 | 0.1393 | 0.0484 | 0.0515 |
| $Q_{C53}$ | 0.0381 | 0.0191 | 0.0529 | 0.1733 |
| $P_{LOSS}$(MW) | 29.20 | 34.0982 | 30.7520 | 31.6099 |
| TVD (p.u) | 0.6501 | 0.6399 | 0.6368 | **0.6311** |
| L-index(p.u) | - | 0.1046 | 0.1549 | 0.1545 |

TABLE 12.
STATISTICAL RESULTS FOR IEEE 57-BUS TEST SYSTEM WITH THE MINIMIZATION OF TVD OBJECTIVE FUNCTION.

| Algorithm | Best (p.u.) | Worst (p.u.) | Mean (p.u.) | Std | % TVD Improve |
|---|---|---|---|---|---|
| OGSA [8] | 0.6982 | - | - | - | 43.40 |
| NGBWCA [87] | 0.6501 | - | - | - | 47.29 |
| ALC-PSO [111] | 0.6634 | 0.6689 | 0.6636 | $89 \times 10^{-5}$ | 46.22 |
| BA [7] | 0.6434 | 0.6609 | 0.6499 | 0.0045 | 47.84 |
| CBA-III [7] | 0.6413 | 0.6542 | 0.6440 | 0.0031 | 48.01 |
| CBA-IV [7] | 0.6399 | 0.6516 | 0.6424 | 0.0028 | 48.13 |
| GBBBA | 0.6368 | 0.6415 | 0.640628 | 0.000706 | 48.375 |
| DeGBBBA | **0.6311** | 0.6397 | 0.636752 | 0.000915 | **48.838** |

The statistical results obtained by different methods are presented in Table 12. It is observed from this table that the DeGBBA can improve the TVD value by 48.838% with respect to initial TVD value, against 48.375% with GBBBA, 48.13% with CBA-IV [7], 48.01% with CBA-III [7], 47.84% with BA [7], 47.29% with NGBWCA [127], 46.22% with ALC-PSO [111] and 43.40% with OGSA [8]. Fig. 6 illustrates the comparison of the convergence characteristics for TVD achieved by DeGBBBA along with the other methods.

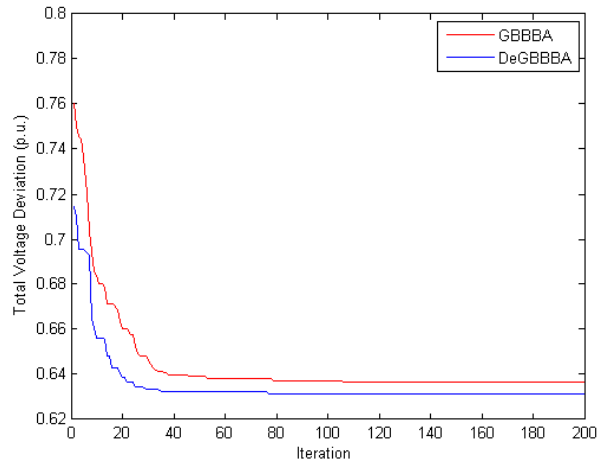

Fig. 6. The convergence characteristics of the GBBBA and DeGBBBA for IEEE 57-bus test power system with the minimization of TVD objective function.

*3) Case 3: Improvement of Voltage Stability Index*

The best results of minimization of the VSI for IEEE 57-bus system by utilizing the proposed algorithms are provided in Table 13. From this table, it is found that the DeGBBBA is better the other algorithms compared with it, because the DeGBBBA achieves an $L-$ index value of 0.157208 in

comparison to 0.157791 with GBBBA, 0.1608 with CBA-IV [7], 0.1789 with CBA-III [7], 0.19001with OGSA [8], 0.190709 with SOA [47] and 0.1917 with BA [7]. The statistical results obtained by different algorithms for the present case are reported in Table 14. In addition, Fig. 7 depicts the comparison of the convergence characteristics for VSI obtained by the proposed algorithms. It is acknowledged from this figure that the DeGBBBA convergence characteristic is better than that obtained by the GBBBA.

TABLE 13.
OPTIMAL SETTINGS OF CONTROL VARIABLES FOR IEEE 57-BUS TEST SYSTEMS WITH THE MINIMIZATION OF VSI OBJECTIVE FUNCTION.

| Variable | CBA-IV [7] | GBBBA | DeGBBBA | Variable | CBA-IV [7] | GBBBA | DeGBBBA |
|---|---|---|---|---|---|---|---|
| Generator voltage (p.u) | | | | Transformer tap ratio (p.u) | | | |
| $V_{G1}$ | 1.0981 | 1.0436 | 0.9326 | $T_{14-46}$ | 1.0253 | 0.9116 | 0.9674 |
| $V_{G2}$ | 1.0441 | 0.9283 | 0.9073 | $T_{10-51}$ | 0.9611 | 1.0146 | 0.9534 |
| $V_{G3}$ | 1.0859 | 1.0471 | 0.9898 | $T_{13-49}$ | 1.0233 | 1.0997 | 0.9857 |
| $V_{G6}$ | 1.0235 | 1.0252 | 1.0210 | $T_{11-43}$ | 0.9150 | 0.9322 | 0.9850 |
| $V_{G8}$ | 1.0607 | 0.9895 | 1.0088 | $T_{40-56}$ | 0.9122 | 0.9635 | 1.0066 |
| $V_{G9}$ | 0.9399 | 0.9538 | 0.9423 | $T_{39-57}$ | 0.9790 | 0.9544 | 1.0580 |
| $V_{G12}$ | 0.9541 | 1.0600 | 1.0509 | $T_{9-55}$ | 0.9426 | 1.0965 | 0.9520 |
| Transformer tap ratio (p.u) | | | | Capacitor banks (p.u) | | | |
| $T_{4-18}$ | 1.0466 | 1.0937 | 0.9037 | $Q_{C18}$ | 0.1784 | 0.1503 | 0.1682 |
| $T_{4-18}$ | 1.0263 | 0.9149 | 1.0005 | $Q_{C25}$ | 0.0391 | 0.1226 | 0.0375 |
| $T_{21-20}$ | 0.9607 | 0.9393 | 0.9450 | $Q_{C53}$ | 0.1714 | 0.0500 | 0.0543 |
| $T_{24-26}$ | 0.9087 | 0.9048 | 0.9753 | $P_{LOSS}$(MW) | 40.3425 | 39.3860 | 39.2251 |
| $T_{7-29}$ | 0.9013 | 0.9491 | 1.0873 | TVD (p.u) | 1.2347 | 1.3273 | 1.3319 |
| $T_{34-32}$ | 1.0804 | 1.0313 | 0.9160 | L-index(p.u) | 0.1608 | 0.157791 | **0.157208** |
| $T_{11-41}$ | 1.0106 | 1.0757 | 1.0417 | | | | |
| $T_{15-45}$ | 0.9982 | 1.0888 | 1.0632 | | | | |

TABLE 14.
STATISTICAL RESULTS FOR IEEE 57-BUS TEST SYSTEM WITH THE MINIMIZATION OF VSI OBJECTIVE FUNCTION.

| Algorithm | Best (p.u.) | Worst (p.u.) | Mean (p.u.) | Std |
|---|---|---|---|---|
| OGSA [8] | 0.190010 | 0.175935 | 0.184112 | $2.8188 \times 10^{-3}$ |
| SOA [47] | 0.190709 | 0.176374 | 0.187451 | $2.6388 \times 10^{-3}$ |
| BA [7] | 0.1917 | 0.2154 | 0.1991 | 0.0075 |
| CBA-III [7] | 0.1789 | 0.1898 | 0.1851 | 0.0035 |
| CBA-IV [7] | 0.1608 | 0.1712 | 0.1674 | 0.0027 |
| GBBBA | 0.157791 | 0.167587 | 0.163586 | 0.002856 |
| DeGBBBA | **0.157208** | 0.166812 | 0.163587 | 0.002998 |

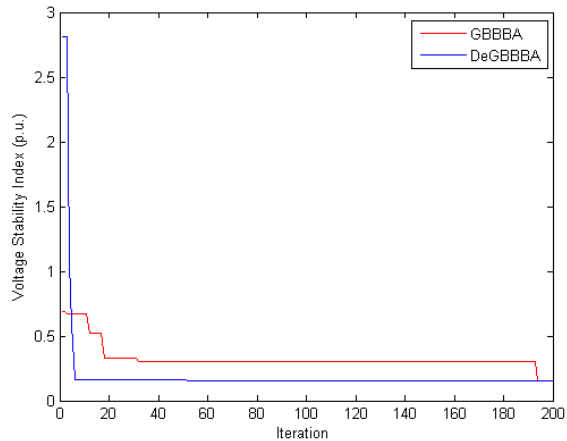

Fig. 7. The convergence characteristics of the GBBBA and DeGBBBA for IEEE 57-bus test power system with the minimization of VSI objective function.

### C. Test System 3: IEEE 118-Bus Test System

The IEEE 118-bus test system is considered as the test system 3. The IEEE 118 bus test system contains 54 generators, 186 transmission lines, 9 tap changing transformers and 14 shunt VAR compensators. Totally, the IEEE 118-bus test system has 77 control variables. The detailed data of this test systems are provided in [47], [53], [101] and [102].

The total demands of the system are: $P_{load} = 4242 MW$ (active power demand), $Q_{load} = 1438 MVAR$ (reactive power demand). The initial total generations and power losses are given by: $\sum PG = 4375.36 MW$ (active power of generators), $\sum QG = 881.92 MVAR$ (reactive power of generators), $P_{LOSS} = 133.357 MW$ (active power losses), $Q_{LOSS} = -785.11 MVAR$ (reactive power losses).

The control variable limits in p.u. are shown in Table 15 [47], [102].

TABLE 15.
CONTROL VARIABLE LIMITS FOR IEEE 118-BUS TEST SYSTEMS [47], [102]

| $V_G^{max}$ | $V_G^{min}$ | $V_{PQ}^{max}$ | $V_{PQ}^{min}$ | $T_k^{max}$ | $T_k^{min}$ | $Q_C^{max}$ | $Q_C^{min}$ |
|---|---|---|---|---|---|---|---|
| 1.06 | 0.94 | 1.1 | 0.95 | 1.1 | 0.9 | 0.2 | 0 |

*1) Case 1: Minimization of Active Power Loss*

The best results yielded by different algorithms for the minimization of the active power loss are depicted in Table 16. From this Table, the DeGBBBA algorithm yields **112.2155 MW** active power loss which outperforms the active power losses yielded by the other algorithms compared with it. The comparative statistical results are tabulated in Table 17. According to this Table, a **15.853%** reduction (from the initial loss of 133.357 MW) in active power loss is obtained achieved by the DeGBBBA algorithm, which outperforms the other algorithms compared with it. The comparative convergence characteristics of active power loss over 300 iterations yielded by the proposed algorithms are depicted in Fig. 8. According to this Figure, the DeGBBBA provides better convergence characteristics than the GBBBA.

TABLE 16.
OPTIMAL SETTINGS OF CONTROL VARIABLES FOR IEEE 118-BUS TEST SYSTEMS WITH THE MINIMIZATION OF $P_{LOSS}$ OBJECTIVE FUNCTION.

| Variable | CBA-IV [7] | GBBBA | DeGBBBA | Variable | CBA-IV [7] | GBBBA | DeGBBBA |
|---|---|---|---|---|---|---|---|
| Generator voltage (p.u) | | | | Generator voltage (p.u) | | | |
| $V_{G1}$ | 0.9810 | 0.9813 | 0.9732 | $V_{G65}$ | 1.0744 | 0.9433 | 1.0464 |
| $V_{G4}$ | 1.0286 | 1.0350 | 1.0218 | $V_{G66}$ | 1.0925 | 1.0243 | 1.0731 |
| $V_{G6}$ | 1.0006 | 1.0524 | 0.9984 | $V_{G69}$ | 1.0701 | 1.0426 | 1.0097 |
| $V_{G8}$ | 1.0995 | 1.0494 | 1.0679 | $V_{G70}$ | 1.0246 | 1.0908 | 1.0083 |
| $V_{G10}$ | 1.0826 | 1.0399 | 1.0838 | $V_{G72}$ | 1.0181 | 1.0901 | 0.9931 |
| $V_{G12}$ | 1.0155 | 1.0458 | 1.0440 | $V_{G73}$ | 1.0147 | 1.0924 | 1.0276 |
| $V_{G15}$ | 1.0245 | 1.0034 | 1.0161 | $V_{G74}$ | 1.0184 | 1.0428 | 0.9979 |
| $V_{G18}$ | 1.0217 | 0.9580 | 0.9806 | $V_{G76}$ | 1.0158 | 1.0094 | 1.0057 |
| $V_{G19}$ | 1.0180 | 0.9975 | 1.0175 | $V_{G77}$ | 1.0579 | 1.0442 | 1.0292 |
| $V_{G24}$ | 1.0281 | 1.0059 | 0.9980 | $V_{G80}$ | 1.0753 | 1.0655 | 1.0489 |
| $V_{G25}$ | 1.0823 | 1.0388 | 1.0636 | $V_{G85}$ | 1.0885 | 1.0835 | 1.0330 |
| $V_{G26}$ | 1.0991 | 1.0163 | 1.0377 | $V_{G87}$ | 1.0730 | 1.0126 | 1.0396 |
| $V_{G27}$ | 1.0404 | 0.9930 | 1.0054 | $V_{G89}$ | 1.1051 | 1.0384 | 1.0561 |
| $V_{G31}$ | 1.0182 | 0.9853 | 1.0214 | $V_{G90}$ | 1.0807 | 0.9859 | 1.0525 |
| $V_{G32}$ | 1.0303 | 1.0316 | 0.9761 | $V_{G91}$ | 1.0896 | 0.9580 | 1.0416 |
| $V_{G34}$ | 1.0446 | 1.0075 | 1.0087 | $V_{G92}$ | 1.1031 | 1.0557 | 1.0539 |
| $V_{G36}$ | 1.0411 | 1.0623 | 1.0200 | $V_{G99}$ | 1.0654 | 0.9759 | 0.9951 |
| $V_{G40}$ | 1.0053 | 1.0148 | 1.0311 | $V_{G100}$ | 1.0804 | 1.0800 | 1.0449 |
| $V_{G42}$ | 1.0090 | 0.9765 | 1.0332 | $V_{G103}$ | 1.0632 | 1.0492 | 1.0289 |
| $V_{G46}$ | 1.0414 | 1.0451 | 1.0055 | $V_{G104}$ | 1.0503 | 1.0287 | 0.9769 |
| $V_{G49}$ | 1.0701 | 1.0148 | 1.0316 | $V_{G105}$ | 1.0546 | 1.0097 | 1.0124 |
| $V_{G54}$ | 1.0462 | 1.0889 | 1.0354 | $V_{G107}$ | 1.0229 | 1.0161 | 0.9953 |
| $V_{G55}$ | 1.0460 | 1.0482 | 1.0028 | $V_{G110}$ | 1.0455 | 0.9996 | 0.9965 |
| $V_{G56}$ | 1.0445 | 0.9670 | 1.0056 | $V_{G111}$ | 1.0415 | 1.0020 | 1.0012 |
| $V_{G59}$ | 1.0740 | 1.0133 | 1.0544 | $V_{G112}$ | 1.0052 | 0.9803 | 0.9636 |
| $V_{G61}$ | 1.0620 | 1.0537 | 1.0554 | $V_{G113}$ | 1.0277 | 1.1806 | 1.0338 |
| $V_{G62}$ | 1.0655 | 1.0155 | 1.0092 | $V_{G116}$ | 1.0764 | 1.0256 | 1.0565 |
| Transformer tap ratio (p.u) | | | | Capacitor banks (p.u) | | | |
| $T_{5-8}$ | 1.0557 | 1.0307 | 0.9773 | $Q_{C46}$ | 0.0700 | 0.1291 | 0.0902 |
| $T_{25-26}$ | 0.9069 | 1.0827 | 0.9863 | $Q_{C48}$ | 0.1687 | 0.0210 | 0.1610 |
| $T_{17-30}$ | 0.9964 | 1.0610 | 0.9981 | $Q_{C74}$ | 0.899 | 0.1079 | 0.1469 |
| $T_{37-38}$ | 1.0062 | 1.0572 | 0.9775 | $Q_{C79}$ | 0.0975 | 0.0585 | 0.1886 |
| $T_{59-63}$ | 0.9642 | 0.9949 | 0.9558 | $Q_{C82}$ | 0.0569 | 0.1924 | 0.1788 |
| $T_{61-64}$ | 1.0350 | 1.0027 | 0.9951 | $Q_{C83}$ | 0.0748 | 0.0265 | 0.0472 |
| $T_{65-66}$ | 1.0328 | 1.0606 | 0.9909 | $Q_{C105}$ | 0.0808 | 0.1551 | 0.0636 |
| $T_{68-69}$ | 0.9568 | 1.1478 | 0.9752 | $Q_{C107}$ | 0.1885 | 0.1670 | 0.0988 |
| $T_{80-81}$ | 0.9515 | 0.9140 | 0.9854 | $Q_{C110}$ | 0.0745 | 0.0839 | 0.0708 |
| Capacitor banks (p.u) | | | | | | | |
| $Q_{C5}$ | 0.1641 | 0.1053 | 0.1338 | $P_{LOSS}$(MW) | 113.7040 | 112.9617 | **112.2155** |
| $Q_{C34}$ | 0.1155 | 0.1327 | 0.1037 | | | | |
| $Q_{C37}$ | 0.1031 | 0.0276 | 0.0883 | TVD (p.u) | 1.5726 | 1.6718 | 1.6098 |
| $Q_{C44}$ | 0.0396 | 0.0332 | 0.0325 | | | | |
| $Q_{C45}$ | 0.1811 | 0.0190 | 0.1252 | L-index(p.u) | 0.0969 | 0.1163 | 0.0969 |

TABLE 17.
STATISTICAL RESULTS FOR IEEE 118-BUS TEST SYSTEM WITH THE MINIMIZATION OF $P_{LOSS}$. OBJECTIVE FUNCTION.

| Algorithm | Best (MW) | Worst (MW) | Mean (MW) | Std | % Psave |
|---|---|---|---|---|---|
| HFA [1] | 134.24 | 134.8499 | 134.96 | 0.008814 | - |

| Algorithm | Best | Worst | Mean | Std | Time |
|---|---|---|---|---|---|
| CLPSO [114] | 130.96 | 132.74 | 131.15 | $85 \times 10^{-6}$ | 1.8 |
| GSA [113] | 127.7603 | - | - | - | 4.2 |
| OGSA [8] | 126.99 | 131.99 | 127.14 | $88 \times 10^{-6}$ | 4.4203 |
| EMA [5] | 126.2243 | 130.98 | 127.0111 | $87.2 \times 10^{-6}$ | 5.35 |
| ALC-PSO [111] | 121.53 | 132.99 | 123.14 | $91 \times 10^{-6}$ | 8.245 |
| GWO [52] | 120.65 | - | - | - | 9.53 |
| ALO [66] | 119.7792 | - | - | - | 9.847 |
| BA [7] | 116.9329 | 119.2345 | 117.1409 | 0.2287 | 12.32 |
| FAHCL-PSO [3] | 116.2479 | - | - | - | 12.83 |
| CBA-III [7] | 115.5989 | 116.3561 | 115.6542 | 0.1569 | 13.32 |
| MICA-IWO [53] | 114.04568 | 114.97562 | 114.44837 | $2.4288 \times 10^{-4}$ | 14.48 |
| MTLA-DDE [102] | 113.9814 | 114.4975 | 114.0852 | $2.8755 \times 10^{-4}$ | 14.53 |
| CBA-IV [7] | 113.7040 | 114.1689 | 114.0108 | 0.1218 | 14.74 |
| GBBBA | 112.9617 | 114.8246 | 113.2743 | 0.22743 | 15.294 |
| QOTLBO [21] | 112.2789 | 115.4516 | 113.7693 | 0.0244 | 15.8 |
| DeGBBBA | **112.2155** | 113.8224 | 112.2505 | 0.226847 | **15.853** |

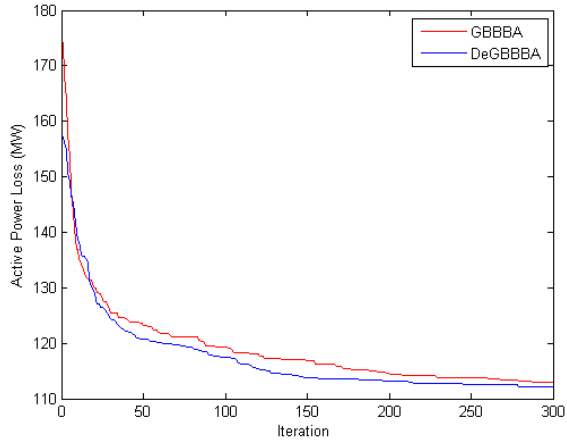

Fig. 8. The convergence characteristics of the GBBBA and DeGBBBA for IEEE 118-bus test power system with the minimization of $P_{LOSS}$ objective function.

*2) Case 2: Minimization of Total Voltage Deviation*

Table 18 illustrates the best results of TVD minimization for IEEE 118-bus system obtained by the proposed algorithms. It is observed from this table that the DeGBBBA is better than the other algorithms compared with it because the DeGBBBA obtains a TVD value of **0.3013** in comparison to 0.3018 with GBBBA, 0.3032 with CBA-IV [7], 0.3059 with CBA-III [7], 0.3172 with BA [7] and 0.3262 with ALC-PSO [111]. The statistical results of this case are reported in Table 19. It is observed from this table that DeGBBBA yields the smallest Best and Mean values compared to other algorithms.

TABLE 18.
OPTIMAL SETTINGS OF CONTROL VARIABLES FOR IEEE 118-BUS TEST SYSTEMS WITH THE MINIMIZATION OF TVD OBJECTIVE FUNCTION.

| Variable | CBA-IV [7] | GBBBA | DeGBBBA | Variable | CBA-IV [7] | GBBBA | DeGBBBA |
|---|---|---|---|---|---|---|---|
| Generator voltage (p.u) | | | | Generator voltage (p.u) | | | |
| $V_{G1}$ | 1.0001 | 0.9925 | 0.9740 | $V_{G65}$ | 0.9838 | 1.0688 | 1.0116 |
| $V_{G4}$ | 1.0000 | 1.0010 | 1.0313 | $V_{G66}$ | 1.0368 | 1.0544 | 1.0158 |
| $V_{G6}$ | 0.9999 | 0.9920 | 1.0083 | $V_{G69}$ | 0.9659 | 0.9758 | 0.9787 |
| $V_{G8}$ | 1.0000 | 0.9997 | 0.9835 | $V_{G70}$ | 1.0055 | 0.9719 | 0.9680 |
| $V_{G10}$ | 1.0002 | 0.9965 | 0.9952 | $V_{G72}$ | 1.0143 | 1.0542 | 1.0244 |
| $V_{G12}$ | 1.0000 | 1.0022 | 1.0212 | $V_{G73}$ | 0.9803 | 1.0414 | 1.0367 |
| $V_{G15}$ | 1.0000 | 1.0044 | 0.9764 | $V_{G74}$ | 1.0076 | 0.9572 | 0.9908 |
| $V_{G18}$ | 1.0000 | 0.9969 | 1.0119 | $V_{G76}$ | 1.0152 | 1.0502 | 1.0062 |
| $V_{G19}$ | 0.9999 | 1.0012 | 1.0170 | $V_{G77}$ | 0.9929 | 1.0032 | 0.9886 |
| $V_{G24}$ | 1.0001 | 0.9991 | 0.9736 | $V_{G80}$ | 1.0108 | 0.9676 | 1.0317 |
| $V_{G25}$ | 0.9999 | 1.0073 | 1.0307 | $V_{G85}$ | 0.9918 | 0.9996 | 0.9539 |
| $V_{G26}$ | 1.0001 | 1.0060 | 1.0268 | $V_{G87}$ | 1.0292 | 0.9867 | 1.0142 |
| $V_{G27}$ | 1.0001 | 1.0146 | 0.9964 | $V_{G89}$ | 1.0435 | 0.9948 | 0.9774 |
| $V_{G31}$ | 1.0000 | 1.0009 | 1.0143 | $V_{G90}$ | 1.0148 | 1.0074 | 0.9710 |
| $V_{G32}$ | 0.9998 | 1.0017 | 0.9795 | $V_{G91}$ | 1.0357 | 1.0281 | 0.9635 |
| $V_{G34}$ | 1.0003 | 1.0010 | 1.0023 | $V_{G92}$ | 0.9963 | 0.9664 | 0.9827 |
| $V_{G36}$ | 1.0000 | 0.9972 | 1.0197 | $V_{G99}$ | 1.0097 | 1.0024 | 0.9876 |
| $V_{G40}$ | 1.0000 | 1.0055 | 0.9978 | $V_{G100}$ | 0.9835 | 0.9516 | 1.0033 |
| $V_{G42}$ | 1.0002 | 0.9969 | 0.9874 | $V_{G103}$ | 0.9771 | 0.9799 | 1.0157 |
| $V_{G46}$ | 1.0000 | 0.9991 | 1.0256 | $V_{G104}$ | 1.0016 | 0.9530 | 1.0011 |
| $V_{G49}$ | 0.9999 | 1.0013 | 1.0125 | $V_{G105}$ | 1.0226 | 1.0224 | 0.9700 |
| $V_{G54}$ | 1.0002 | 0.9980 | 0.9966 | $V_{G107}$ | 1.0270 | 1.0592 | 1.0043 |
| $V_{G55}$ | 0.9997 | 0.9952 | 0.9799 | $V_{G110}$ | 0.9864 | 1.0552 | 1.0377 |
| $V_{G56}$ | 1.0001 | 1.0028 | 1.0100 | $V_{G111}$ | 0.9744 | 0.9617 | 1.0100 |
| $V_{G59}$ | 0.9999 | 0.9940 | 0.9795 | $V_{G112}$ | 0.9861 | 1.0140 | 1.0305 |
| $V_{G61}$ | 1.0000 | 0.9991 | 1.0116 | $V_{G113}$ | 1.0189 | 0.9481 | 0.9845 |
| $V_{G62}$ | 0.9999 | 1.0004 | 0.9852 | $V_{G116}$ | 1.0289 | 1.0089 | 0.9862 |
| Transformer tap ratio (p.u) | | | | Capacitor banks (p.u) | | | |
| $T_{5-8}$ | 1.0441 | 1.0718 | 0.9683 | $Q_{C46}$ | 0.1283 | 0.1237 | 0.0664 |
| $T_{25-26}$ | 0.9630 | 0.9346 | 0.9999 | $Q_{C48}$ | 0.1448 | 0.1678 | 0.0372 |
| $T_{17-30}$ | 1.0094 | 1.0835 | 1.0400 | $Q_{C74}$ | 0.0810 | 0.1185 | 0.1470 |
| $T_{37-38}$ | 1.0776 | 1.0721 | 0.9373 | $Q_{C79}$ | 0.1032 | 0.1814 | 0.0209 |
| $T_{59-63}$ | 0.9270 | 0.9331 | 1.0797 | $Q_{C82}$ | 0.0709 | 0.0290 | 0.1342 |
| $T_{61-64}$ | 1.0492 | 1.0918 | 1.0013 | $Q_{C83}$ | 0.0520 | 0.1260 | 0.1094 |
| $T_{65-66}$ | 1.0400 | 0.9450 | 1.0300 | $Q_{C105}$ | 0.1013 | 0.1991 | 0.1583 |
| $T_{68-69}$ | 0.9521 | 1.0387 | 1.0240 | $Q_{C107}$ | 0.1642 | 0.0094 | 0.1117 |
| $T_{80-81}$ | 0.9729 | 1.0732 | 0.9731 | $Q_{C110}$ | 0.0719 | 0.0078 | 0.1340 |
| Capacitor banks (p.u) | | | | $P_{LOSS}$(MW) | 142.1661 | 141.4920 | 144.0433 |
| $Q_{C5}$ | 0.0730 | 0.1232 | 0.0664 | | | | |
| $Q_{C34}$ | 0.1303 | 0.0763 | 0.1410 | TVD (p.u) | 0.3032 | 0.3018 | **0.3013** |
| $Q_{C37}$ | 0.1174 | 0.0320 | 0.1227 | | | | |
| $Q_{C44}$ | 0.0902 | 0.1362 | 0.1636 | L-index(p.u) | 0.1189 | 0.1357 | 0.1231 |
| $Q_{C45}$ | 0.1228 | 0.1209 | 0.1304 | | | | |

TABLE 19.
STATISTICAL RESULTS FOR IEEE 118-BUS TEST SYSTEM WITH THE MINIMIZATION OF TVD OBJECTIVE FUNCTION.

| Algorithm | Best (p.u.) | Worst (p.u.) | Mean (p.u.) | Std |
|---|---|---|---|---|
| ALC-PSO [111] | 0.3262 | 0.3743 | 0.3281 | $95 \times 10^{-6}$ |
| BA [7] | 0.3172 | 0.3206 | 0.3187 | $9.7474 \times 10^{-4}$ |
| CBA-III [7] | 0.3059 | 0.3072 | 0.3062 | $3.2319 \times 10^{-4}$ |
| CBA-IV [7] | 0.3032 | 0.3041 | 0.3036 | $3.0558 \times 10^{-4}$ |
| GBBBA | 0.3018 | 0.3030 | 0.302736 | 0.00021 |
| DeGBBBA | **0.3013** | 0.3027 | 0.302408 | 0.000209 |

Fig. 9 shows the comparison of the convergence characteristics for TVD achieved by DeGBBBA with the other methods and also this figure reveals shows the optimization capability of the DeGBBBA for larger dimension systems.

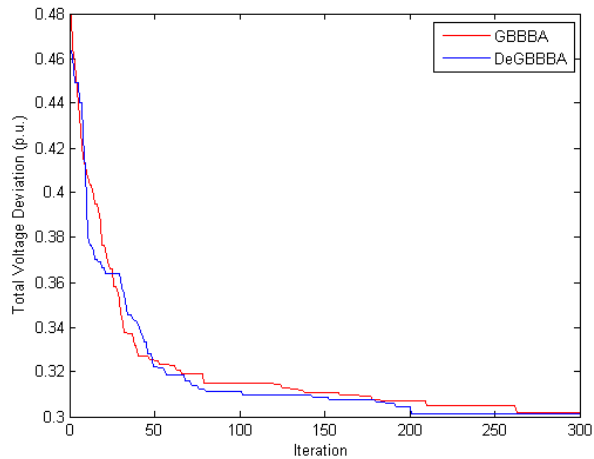

Fig. 9. The convergence characteristics of the GBBBA and DeGBBBA for IEEE 118-bus test power system with the minimization of TVD objective function.

*3) Case 3: Improvement of Voltage Stability Index*

Table 20 shows the best results of minimization of the VSI for IEEE 118-bus system obtained by employing the DeGBBBA algorithm. According to this table, the DeGBBBA is better than the other algorithms compared with it because the DeGBBBA gives an $L-$ index value of **0.0543** in comparison to 0.0558 with GBBBA, 0.0570 with CBA-IV [78], 0.0587 with CBA-III [78], 0.0608 with QOTLBO [21] and 0.06126 with BA [7]. For the sake of comparison, Table 21 summarizes the statistical results for the present case. The comparison of the convergence characteristics for VSI achieved by the proposed algorithms is shown in Fig. 10 and demonstrates the robustness of the DeGBBBA in solving larger dimension systems.

TABLE 20.
OPTIMAL SETTINGS OF CONTROL VARIABLES FOR IEEE 118-BUS TEST SYSTEMS WITH THE MINIMIZATION OF VSI OBJECTIVE FUNCTION.

| Variable | CBA-IV [7] | GBBBA | DeGBBBA | Variable | CBA-IV [7] | GBBBA | DeGBBBA |
|---|---|---|---|---|---|---|---|
| Generator voltage (p.u) | | | | Generator voltage (p.u) | | | |
| $V_{G1}$ | 1.0307 | 0.9620 | 0.9726 | $V_{G65}$ | 1.0009 | 0.9800 | 1.0205 |
| $V_{G4}$ | 1.0197 | 0.9949 | 1.0110 | $V_{G66}$ | 1.0045 | 1.0720 | 0.9728 |
| $V_{G6}$ | 1.0018 | 0.9949 | 1.0225 | $V_{G69}$ | 0.9753 | 1.0688 | 1.0055 |
| $V_{G8}$ | 1.0379 | 0.9728 | 1.0197 | $V_{G70}$ | 1.0415 | 1.0700 | 0.9895 |
| $V_{G10}$ | 0.9986 | 0.9527 | 1.0293 | $V_{G72}$ | 0.9646 | 1.0783 | 1.0096 |
| $V_{G12}$ | 1.0097 | 1.0564 | 1.0086 | $V_{G73}$ | 1.0118 | 1.0055 | 1.0913 |
| $V_{G15}$ | 1.0074 | 1.0139 | 1.0163 | $V_{G74}$ | 1.0159 | 0.9723 | 1.0258 |
| $V_{G18}$ | 0.9739 | 1.0006 | 0.9566 | $V_{G76}$ | 0.9907 | 1.0849 | 0.9669 |
| $V_{G19}$ | 0.9679 | 1.0347 | 1.0444 | $V_{G77}$ | 0.9541 | 1.0087 | 1.0732 |
| $V_{G24}$ | 1.0131 | 0.9646 | 0.9644 | $V_{G80}$ | 0.9640 | 1.0033 | 1.0582 |
| $V_{G25}$ | 1.0157 | 1.0071 | 0.9876 | $V_{G85}$ | 0.9645 | 1.0133 | 0.9567 |
| $V_{G26}$ | 1.0337 | 1.0329 | 1.0243 | $V_{G87}$ | 0.9907 | 1.0443 | 1.0363 |
| $V_{G27}$ | 1.0021 | 0.9633 | 0.9521 | $V_{G89}$ | 1.0053 | 1.0476 | 0.9948 |
| $V_{G31}$ | 1.0539 | 0.9899 | 0.9615 | $V_{G90}$ | 0.9952 | 0.9931 | 0.9951 |
| $V_{G32}$ | 0.9627 | 0.9665 | 0.9868 | $V_{G91}$ | 0.9560 | 0.9551 | 0.9951 |
| $V_{G34}$ | 0.9711 | 0.9936 | 0.9909 | $V_{G92}$ | 1.0031 | 0.9955 | 1.0261 |
| $V_{G36}$ | 1.0185 | 0.9676 | 1.0586 | $V_{G99}$ | 1.0058 | 0.9688 | 1.0375 |
| $V_{G40}$ | 1.0371 | 1.0020 | 0.9789 | $V_{G100}$ | 1.0212 | 0.9920 | 0.9945 |
| $V_{G42}$ | 1.0017 | 1.0208 | 1.0200 | $V_{G103}$ | 1.0221 | 0.9745 | 1.0066 |
| $V_{G46}$ | 0.9504 | 1.0320 | 1.0278 | $V_{G104}$ | 1.0107 | 0.9506 | 0.9650 |
| $V_{G49}$ | 0.9996 | 0.9545 | 0.9731 | $V_{G105}$ | 1.0156 | 1.0564 | 0.9679 |
| $V_{G54}$ | 1.0300 | 0.9712 | 1.0139 | $V_{G107}$ | 1.0488 | 0.9596 | 1.0092 |
| $V_{G55}$ | 0.9785 | 0.9896 | 1.0672 | $V_{G110}$ | 1.0390 | 1.0228 | 1.0339 |
| $V_{G56}$ | 1.0294 | 0.9679 | 1.0474 | $V_{G111}$ | 0.9646 | 1.0303 | 0.9553 |
| $V_{G59}$ | 0.9918 | 1.0034 | 1.0298 | $V_{G112}$ | 0.9859 | 1.0184 | 1.0523 |
| $V_{G61}$ | 0.9987 | 0.9737 | 0.9917 | $V_{G113}$ | 1.0088 | 0.9858 | 0.9696 |
| $V_{G62}$ | 0.9787 | 0.9475 | 0.9782 | $V_{G116}$ | 1.0102 | 0.9524 | 0.9947 |
| Transformer tap ratio (p.u) | | | | Capacitor banks (p.u) | | | |
| $T_{5-8}$ | 0.9633 | 1.0381 | 0.9619 | $Q_{C46}$ | 0.1321 | 0.1232 | 0.0750 |
| $T_{25-26}$ | 0.9623 | 1.0843 | 0.9694 | $Q_{C48}$ | 0.1125 | 0.1908 | 0.0332 |
| $T_{17-30}$ | 0.9839 | 0.9302 | 0.9641 | $Q_{C74}$ | 0.1706 | 0.0396 | 0.1114 |
| $T_{37-38}$ | 1.0315 | 1.0597 | 0.9596 | $Q_{C79}$ | 0.0539 | 0.0968 | 0.0432 |
| $T_{59-63}$ | 0.9762 | 0.9172 | 0.9427 | $Q_{C82}$ | 0.0631 | 0.0566 | 0.1813 |
| $T_{61-64}$ | 0.9681 | 1.0006 | 0.9481 | $Q_{C83}$ | 0.0723 | 0.1762 | 0.0615 |
| $T_{65-66}$ | 1.0103 | 1.0658 | 1.0415 | $Q_{C105}$ | 0.1268 | 0.0329 | 0.1580 |
| $T_{68-69}$ | 0.9323 | 0.9518 | 0.9951 | $Q_{C107}$ | 0.1564 | 0.1086 | 0.1446 |
| $T_{80-81}$ | 1.0543 | 1.0335 | 0.9086 | $Q_{C110}$ | 0.0573 | 0.1208 | 0.0078 |
| Capacitor banks (p.u) | | | | $P_{LOSS}$(MW) | 169.5624 | 174.3956 | 183.4976 |
| $Q_{C5}$ | 0.1155 | 0.0537 | 0.0773 | | | | |
| $Q_{C34}$ | 0.1551 | 0.1090 | 0.1774 | $TVD$ (p.u) | 1.4538 | 1.5966 | 1.6668 |
| $Q_{C37}$ | 0.1724 | 0.0801 | 0.0072 | | | | |
| $Q_{C44}$ | 0.0786 | 0.1793 | 0.0748 | L-index(p.u) | 0.0570 | 0.0558 | **0.0543** |
| $Q_{C45}$ | 0.1032 | 0.0430 | 0.1348 | | | | |

TABLE 21.
STATISTICAL RESULTS FOR IEEE 118-BUS TEST SYSTEM WITH THE MINIMIZATION OF VSI OBJECTIVE FUNCTION.

| Algorithm | Best (p.u.) | Worst (p.u.) | Mean (p.u.) | Std |
|---|---|---|---|---|
| BA [7] | 0.06126 | 0.0654 | 0.0632 | 0.0010 |
| QOTLBO | 0.0608 | 0.06311 | 0.0616 | 0.0476 |
| CBA-III [7] | 0.0587 | 0.0626 | 0.0607 | 0.0013 |
| CBA-IV [7] | 0.0570 | 0.0609 | 0.0587 | 0.0012 |
| GBBBA | 0.0558 | 0.0582 | 0.057164 | 0.00503 |
| DeGBBBA | **0.0543** | 0.0577 | 0.056858 | 0.000412 |

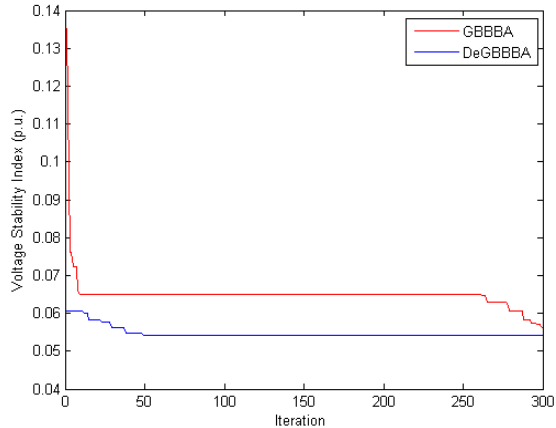

Fig. 10. The convergence characteristics of the GBBBA and DeGBBBA for IEEE 118-bus test power system with the minimization of VSI objective function.

## VI. CONCLUSION

In this paper, two newly developed variants of bat algorithm, named as GBBBA and DeGBBBA, have been successfully implemented for solving the ORPD problem of electric power systems. The proposed algorithms offer advantages over the standard BA and other metaheuristics as they employ the Gaussian distribution and the modified Gaussian distribution to overcome the premature convergence while dealing with more complex problems such as ORPD problem. The proposed GBBBA and DeGBBBA were tested on IEEE 14-bus, 57-bus and 118-bus test power systems to demonstrate their effectiveness and robustness. The results obtained by GBBBA and DeGBBBA algorithms were compared to those from the other algorithms reported in the literature. Simulation results evidently show that the proposed DeGBBBA yields better results than the other algorithms compared with it in handling ORPD problem.

For future work, the authors will apply the proposed algorithms to solve ORPD problem incorporating renewable sources such as (stochastic) wind and solar (PV) power. It would be also interesting to employ some other distributions, including exponential distribution and Cauchy distribution in order, to enhance the optimization capabilities of the standard bat algorithm. It is an interesting topic is to extend the developed improved bat algorithm for the optimal scheduling of integrated energy systems with renewables [115], [116]. It would also be interesting to develop a hybrid analytical-heuristic approach for solving the optimal reactive power dispatch problem of power systems [117].

## REFERENCES

[1] A. Rajan, and T. Malakar, "Optimal reactive power dispatch using hybrid Nelder–Mead simplex based firefly algorithm," *Int. J. Electr.Power Energy Syst.*, vol. 66, pp. 9-24, 2015.
[2] M. Ghasemi, S. Ghavidel, M.M. Ghanbarian, and A. Habibi, "A new hybrid algorithm for optimal reactive power dispatch problem with discrete and continuous control variable," *Applied soft Computing*, vol. 22, pp. 126-140, 2014.
[3] E. Naderi, H. Narimani, M. Fathi, and M. R. Narimani, "A novel fuzzy adaptive configuration of particle swarm optimization to solve large scale optimal reactive power dispatch," *Appl. Soft Comput.*, vol. 53, pp. 441-456, Jan. 2017.
[4] R. Mallipeddi, S. Jeyadevi, P. N. Suganthan, and S.Baskar, "Efficient constraint handling for optimal reactive power dispatch problems," *Swarm Evol. Comput.,* vol.5, pp. 28-36, Aug. 2012.
[5] A. Rajan, and T. Malakar, "Exchange market algorithm based optimum reactive power dispatch", *Appl. Soft Comput.*, vol. 43, pp. 320-336, 2016.
[6] G. Chen, L. Liu, Z. Zhizhong, and H. Shanwai, "Optimal reactive power dispatch by improved GSA-based algorithm with the novel strategies to handle constraints," *Appl. Soft Comput*., vol.50, pp.58-70, Jan. 2017.
[7] S. Mugemanyi, Z. Qu, F. X. Rugema, Y. Dong, C. Bananeza, and L. Wang, "Optimal reactive power dispatch using chaotic bat algorithm," in *IEEE Access*, vol. 8, pp. 65830-65867, 2020.
[8] B. Shaw, V. Mukherjee, and S.P. Ghoshal, "Solution of reactive power dispatch of power systems by an opposition-based gravitational search algorithm," *Int. J. Electr. Power Energy Syst*., vol. 55, pp. 29-40, 2014.
[9] N.I. Deeb, and S.M. Shahidehpour, "An efficient technique for reactive power dispatch using a revised linear programming approach," *Electr. Power Syst. Res*. vol. 15, no. 2, pp. 121-134, 1988.
[10] D. Kirschen, and H. Van Meeteren, "MW/voltage control in a linear programming based optimal power flow," *IEEE Trans. Power Syst.*, vol. 3, no. 2, pp. 481-489, May 1988.
[11] S.S. Sachdeva, and R. Billinton, "Optimum Network VAR Planning by Nonlinear Programming", *IEEE Trans. Power App. Syst.*, vol. P AS-92, pp. 1217-1225, 1973.
[12] F.-C. Lu, and Y.-Y. Hsu, "Reactive power/voltage control in a distribution substation using dynamic programming," *IEE Proc. Gener. Transm. Distrib*., vol. 142, no. 6, pp. 639-645, 1995.
[13] K.L. Lo, and S.P. Zhu, "A decoupled quadratic programming appro ach for optimal power dispatch," *Electric Power System Research*, v ol. 22, no. 1, pp. 47-60, 1991.
[14] N. Grudinin, "Reactive power optimization using successive quadratic programming method," *IEEE Transactions on Power Systems*, vol. 13, no. 4, pp. 1219-1225, 1998.
[15] V.H. Quintana, and M. Santos-Nieto, "Reactive-power dispatch by successive quadratic programming," *IEEE Trans. Energy Convers.*, vol. 4, no. 3, pp. 425-435, 1989.
[16] J.A. Momoh, S.X. Guo, E.C. Oghuobiri, and R. Adapa, "The quadratic interior point method solving the power system optimization problems," *IEEE Trans Power Syst*, vol. 9, no. 3, pp. 1327-1336, 1994.
[17] K. Aoki, M. Fan, and A. Nishikori, "Optimal VAR planning by approximation method for recursive mixed-integer linear programming," *IEEE Trans. Power Syst.*, vol. 3, no. 4, pp. 1741-1747, 1988.

[18] S. Granville, "Optimal reactive dispatch through interior point methods," *IEEE Trans. Power Syst.*, vol. 9, no. 1, pp. 136-146, Feb. 1994.

[19] W. Yan, J. Yu, D. C. Yu, and K. Bhattarai, "A new optimal reactive power flow model in rectangular form and its solution by predictor corrector primal dual interior point method," *IEEE Trans. Power Syst.*, vol. 21, no. 1, pp. 61-67, Feb. 2006.

[20] M. Zhang, and Y. Li. "Multi-objective optimal reactive power dispatch of power systems by combining classification-based Multi-objective evolutionary algorithm and integrated decision making," in *IEEE Access*, vol. 8, pp. 38198-38209, 2020.

[21] B. Mandal, and P. K. Roy, "Optimal reactive power dispatch using quasi-oppositional teaching learning based optimization", *Int. J. Electr. Power Energy Syst.*, vol.53, pp. 123–134, 2013.

[22] M. Mehdinejad, B. Mohammadi-Ivatloo, R. Dadashzadeh-Bonab, and K. Zare, "Solution of optimal reactive power dispatch of power systems using hybrid particle swarm optimization and imperialist competitive algorithms," *Int. J. Electr. Power Energy Syst.*, vol. 83, pp. 104–116, Apr. 2016.

[23] S. Mirjalili, A. H. Gandomi, S. Z. Mirjalili, S. Saremi, H. Faris, and S. M. Mirjalili, "Salp Swarm Algorithm: A bio-inspired optimizer for engineering design problems," *Adv. Eng. Software*, vol. 114, pp.163-191, 2017.

[24] A. A. Heidari, S. Mirjalili, H. Faris, I. Aljarah, M. Mafarja, and H. Chen, "Harris hawks optimization: Algorithm and applications," *Future Generation Computer Systems*, vol. 97, pp. 849-872, 2019.

[25] F. A. Hashim, E. H. Houssein, M. S. Mabrouk, W. Al-Atabany, and S. Mirjalili, "Henry gas solubility optimization: A novel physics-based algorithm," *Future Generation Computer Systems*, vol. 101, pp. 646-667, 2019.

[26] D. Simon, "Biogeography-based optimization," *IEEE Trans. Evol. Comput.*, vol. 12, pp. 702-713, 2008

[27] J.R. Koza, "Genetic Programming II, Automatic Discovery of Reusable Subprograms," *MIT Press*, Cambridge, MA, 1992.

[28] J. H. Holland, "Genetic algorithms and the optimal allocation of trials," *SlAM J. Comput.*, vol. 2, pp. 88-105, 1973.

[29] R. Storn, and K. Price, "Differential evolution: A simple and efficient heuristic for global optimization over continuous spaces," *J. Global Optimization*, vol. 11, no. 4, pp. 341–359, 1997.

[30] X. Yao, L. Liu, and G. Lin, "Evolutionary programming made faster," *Evolut Comput. IEEE Trans*, vol. 3, pp. 82-102, 1999.

[31] I. Rechenberg, "Evolutionsstrategien," in: *Simulationsmethoden in derMedizin und Biologie*, Springer, pp. 83-114, 1978.

[32] S. Kirkpatrick, C.D. Gelatt Jr., and M.P. Vecchi, "Optimization by simulated annealing," *Science*, vol. 220, pp. 671-680, 1983.

[33] E. Rashedi, H.Nezamabadi-Pour, and S. Saryazdi, "GSA: a gravitational search algorithm," *Information Science*, vol. 179, no. 13, pp. 2232-2248, 2009.

[34] S. Mirjalili, and A. Lewis, "The whale optimization algorithm," *Adv. Eng. Software*, vol. 95, pp. 51-67, May 2016.

[35] Z. W. Geem, J. M. Kim, and G. V. Loganathan, "A new heuristic optimization algorithm: harmony search," *Simulation*, vol. 76, no. 2, pp.60-68, 2001.

[36] F. Glover, "Tabu search-Part I," *ORSA J. Comput.*, vol. 1, pp. 190-206, 1989.

[37] F. Glover, "Tabu search-Part II," *ORSA J. Comput.*, vol. 2, pp. 4-32, 1989.

[38] R.V. Rao, V.J. Savsani, and D. Vakharia, "Teaching-learning-based optimization:an optimization method for continuous non-linear large scale problems," *Inform. Sci.*, vol. 183, pp. 1-15, 2012.

[39] E. Atashpaz-Gargari, and C. Lucas, "Imperialist competitive algorithm: An algorithm for optimization inspired by imperialistic competition," in: *2007 IEEE Congr. Evol. Comput*., 2007, pp. 4661-4667, 2007.

[40] C. Dai, Y. Zhu, and W. Chen, "Seeker optimization algorithm," in *Lecture Notes in Artificial Intelligence*, Y. Wang, Y. Cheung, and H. Liu, Eds. Berlin, Germany: Springer-Verlag, 2007, pp. 167-176, CIS 2006.

[41] N. Ghorbani, and E. Babaei, "Exchange market algorithm," *Appl. Soft Comput.*, vol. 19, pp. 177-187, 2014.

[42] J. Kennedy, and R. C. Eberhart, "Particle swarm optimization," *Proc. IEEE Int. Conf. Neural Netw*, vol. 4, pp. 1942-1948, 1995.

[43] M. Dorigo, M. Birattari, and T. Stützle, "Ant colony optimization," *IEEE Comput. Intell. Mag*., vol. 1, no. 4, pp. 28-39, 2006.

[44] X.-S. Yang, and X. He, "Bat algorithm: literature review and applications," *International Journal of Bio-Inspired Computation*, vol. 5, no. 3, pp. 141- 149, 2013.

[45] S. Mirjalili, "Dragonfly algorithm: a new meta-heuristic optimization technique for solving single-objective discrete and multi-objective problems," *Neural Computing and Applications*, vol. 27, no. 4, pp. 1053-1073, 2016.

[46] S. Mirjalili, "Grasshopper optimisation algorithm: theory and application," *Adv. Eng. Software*, vol. 105, pp. 30-47, 2016.

[47] C. Dai, W. Chen, Y. Zhu, and X. Zhang, "Seeker optimization algorithm for optimal reactive power dispatch," *IEEE Trans. Power Syst.*, vol. 24, no. 3, pp. 1218–1231, Aug. 2009.

[48] G. Chen, L. Liu, Z. Zhizhong, and H. Shanwai, "Enhanced GSA-based optimization for minimization of power losses in power system," *Mathematical Problems in Engineering.* vol. 10, pp.1-13, 2015.

[49] M. R. Babu, and M. Lakshmi, "Gravitational search algorithm based approach for optimal reactive power dispatch," *The 2th Int. Conf. on Science Techn. Eng. and Manag. (ICONSTEM)*, pp. 360-366, 2016.

[50] R. N. S. Mei, M. H. Sulaiman, Z. Mustaffa, and H. Daniyal, "Optimal reactive power dispatch solution by loss minimization using moth-flame optimization technique," *Appl. Soft. Comput.*, vol. 59, pp. 210- 222, 2017.

[51] M. Ghasemi, S. Ghavidel, M.M. Ghanbarian, and M. Gitizadeh, "Multi-objective optimal electric power planning in the power system using Gaussian bare-bones imperialist competitive algorithm," *Inf. Sci.*, vol. 294, pp. 286-304, 2015.

[52] M. H. Sulaiman, Z. Mustaffa, M. R. Mohamed, and O. Aliman, "Using the gray wolf optimizer for solving optimal reactive power dispatch problem," *Appl. Soft Comput.*, vol. 32, pp. 286–292, Apr. 2015.

[53] M. R. Nayak, K. R. Krishnanand, and P. K. Rout, "Optimal reactive power dispatch based on adaptive invasive weed optimization algorithm," *IEEE International Conference on Energy Automation and Signal*, pp. 1-7, Dec. 2011.

[54] N. Sinsuphan, U. Leeton, and T. Kulworawanichpong, "Optimal power flow solution using improved harmony search method," *Appl. Soft Comput.*, vol. 13, no. 5, pp. 2364-2374, 2013.

[55] M. Ghasemi, M. Taghizadeh, S. Ghavidel, J. Aghaei, and A. Abbasian, "Solving optimal reactive power dispatch problem using a novel teaching-learning-based optimization algorithm," *Eng. Appl. Artif. Intell.*, vol. 39, no. 3, pp. 100-108, Mar. 2015.

[56] A. Bhattacharya, and P.K. Chattopadhyay, "Solution of optimal reactive power flow using biogeography-based optimization," *Int. J. Energy Power Eng*., vol. 3, pp. 269-277, 2010.

[57] H. Yoshida, K. Kawata, Y. Fukuyama, S. Takayama, and Y. Nakanishi, "A particle swarm optimization for reactive power and voltage control considering voltage security assessment," *IEEE Trans. Power Syst.*, vol. 15, no. 4, pp. 1232–1239, 2000.

[58] A.A.A. Esmin, G. Lambert-Torres, and A.C.Z. de Souza, "A hybrid particle swarm optimization applied to loss power minimization," *IEEE Trans. Power Syst.*, vol. 20, no. 2, pp. 859-866, 2005.

[59] M. H. Sulaiman, and Z. Mustaffa, "Cuckoo search algorithm as an optimizer for optimal reactive power dispatch problems," *Control Automation and Robotics (ICCAR) 2017 3rd International Conference on*, pp. 735-739, 2017.

[60] D. Devaraj, "Improved genetic algorithm for multi-objective reactive power dispatch Problem," *Eur. Trans. Electr. Power*, vol. 17, no. 6, pp. 569-581, 2007.

[61] P. Subbaraj, and P. N. Rajnarayanan, "Optimal reactive power dispatch using self-adaptive real coded genetic algorithm," *Electric Power Systems Research*, vol. 79, pp. 374-381, 2009.

[62] Q. H. Wu, Y. J. Cao, and J. Y. Wen, "Optimal reactive power dispatch using an adaptive genetic algorithm," *Int. J. Electr. Power Energy Syst.*, vol. 20, no. 8, pp. 563-569, 1998.

[63] A. Mukherjee, and V. Mukherjee, "Solution of optimal reactive power dispatch by chaotic krill herd algorithm," *IET Gener Transm Distrib*, vol. 9, no. 15, pp. 2351-2362, 2015.

[64] M. Varadarajan, and K. S. Swarup, "Differential evolution approach for optimal reactive power dispatch," *Applied Soft Computing*, vol. 8, pp. 1549-1561, 2008.

[65] M. Tripathy, and S. Mishra, "Bacteria foraging-based solution to optimize both real power loss and voltage stability limit," *IEEE Trans. Power Syst.*, vol. 22, no. 1, pp. 240-248, 2007.

[66] S. Moussa, T. Bouktir, and A. Salhi, "Ant lion optimizer for solving optimal reactive power dispatch problem in power systems," *Engineering Science and Technology, an International Journal*, vol. 20, no. 3, pp. 885-895, June 2017.

[67] A. A. Heidari, R.. A. Abbaspour, and A. R. Jordehi, "Gaussian bare-bones water cycle algorithm for optimal reactive power dispatch in electrical power systems," *Appl. Soft Comput.*, vol. 57, pp. 657-671, August 2017.

[68] X.-S. Yang, "A new metaheuristic bat-inspired algorithm," *Nature Inspired Cooperative Strategies for Optimization (NICSO 2010)*, vol. 284, pp. 65-74, 2010.

[69] B. Varsha, K. Manoj, and S. Vrajesh, "A survey on swarm intelligence techniques," *International Interdisciplinary Conference on Science Technology Engineering Management Pharmacy and Humanities*, Singapore, pp. 238-246, 22-23 April 2017.

[70] Q. Liu, J. Li, L. Wu, F. Wang, and W. Xiao, "A novel bat algorithm with double mutation operators and its application to low-velocity impact localization problem," *Engineering Applications of Artificial Intelligence* vol. 90, pp. 1-19, 2020.

[71] J. W. Zhang, and G. G. Wang, "Image matching using a bat algorithm with mutation," *Applied Mechanics and Materials*, vol. 203, no. 1, pp. 88-93, 2012.

[72] X.-S. Yang, and A. Hossein Gandomi, "Bat algorithm: a novel approach for global engineering optimization," *Engineering Computations*, vol.29, no.5, pp.464-483, 2012.

[73] A. Kaveh, and P. Zakian, "Enhanced bat algorithm for optimal design of skeletal structures," *Asian Journal of Civil Engineering*, vol. 15, no.2, pp.179-212, 2014.

[74] X.-S. Yang, "Bat algorithm for multi-objective optimisation," *International Journal of Bio-Inspired Computation*, vol.3, no,5, pp.267-274, 2011.

[75] F. X. Rugema, G. Yan, S. Mugemanyi, Q. Jia, H. Zhang, and C. Bananeza, "A Cauchy-Gaussian quantum-behaved bat algorithm applied to solve the economic load dispatch problem," in *IEEE Access*, vol. 9, pp. 3207-3228, 2021.

[76] J. Tholath Jose, "Economic load dispatch including wind power using bat algorithm," *2014 International Conference on Advances in Electrical Engineering (ICAEE)*, pp. 1-4, Vellore, 2014.

[77] B. R. Adarsh, T. Raghunathan, and T. Jayabarathi, "Economic disp atch using chaotic bat algorithm," *Energy*, vol. 96, pp. 666-675, 201 6.

[78] J. H. Lin, C. W. Chou, C. H. Yang, and H. L. Tsai, "A chaotic Lévy flight bat algorithm for parameter estimation in nonlinear dynamic biological systems," *J. Computer and Information Technology*, vol. 2, no. 2, pp. 56–63, 2012.

[79] X.-S. Yang, M. Karamanoglu, and S. Fong, "Bat algorithm for topology optimization in microelectronic applications," *The First International Conference on Future Generation Communication Technologies*, pp. 150-155, London, 2012.

[80] K. Khan, and A. Sahai, "A fuzzy c-means bisonar-based metaheuristic optimization algorithm," *Int. J. of Interactive Multimedia and Artificial Intelligence*, vol. 1, no. 7, pp. 26-32, 2012.

[81] G. Komarasamy, and A. Wahi, "An optimized K-means clustering technique using bat algorithm," *European J Scientific Research*, vol. 84, no. 2, pp. 263-273, 2012.

[82] K. Khan, and A. Sahai, "A comparison of BA, GA, PSO, BP and LM for training feed forward neural networks in e-learning context," *Int. J. Intelligent Systems and Applications (IJISA)*, vol. 4, no. 7, pp. 23-29, 2012.

[83] A. Natarajan, S. Subramanian, and K. Premalatha, "A comparative study of cuckoo search and bat algorithm for bloom filter optimisation in spam filtering," *Int. J. Bio-Inspired Computation*, vol. 4, no. 2, pp. 89-99, 2012.

[84] R. Y. M. Nakamura, L. A. M. Pereira, K. A. Costa, D. Rodrigues, J. P. Papa, and X.-S. Yang, "BBA: A binary bat algorithm for feature selection," in: *25th SIBGRAPI Conference on Graphics, Patterns and Images (SIBGRAPI)*, *IEEE Publication*, pp. 291-297, 22-25 Aug. 2012.

[85] K. Khan, A. Nikov, and A. Sahai, "A fuzzy bat clustering method for ergonomic screening of office workplaces," *Third International Conference on Software Services and Semantic Technologies S3T*, vol. 101, no. 1, pp. 59-66, 2011.

[86] S. Mishra, K. Shaw, and D. Mishra, "A new metaheuristic bat inspired classification approach for microarray data," *Procedia Technology*, vol. 4, no. 1, pp. 802-806, 2012.

[87] R. Damodaram, and M. L. Valarmathi, "Phishing website detection and optimization using modified bat algorithm," *Int. J. Engineering Research and Applications*, vol. 2, no. 1, pp. 870-876, 2012.

[88] P. Musikapun, and P. Pongcharoen, "Solving multi-stage multi-machine multi-product scheduling problem using bat algorithm," *2nd International Conference on Management and Artificial Intelligence (IPEDR)*, vol. 35, IACSIT Press, Singapore, pp. 98–102, 2012.

[89] Kennedy, "Bare-bones particle swarms," in *Proceedings of the IEEE Swarm Intelligence Symposium*, pp. 80-87, 2003.

[90] R. A. Krohling, and E. Mendel, "Bare-bones particle swarm optimization with Gaussian or Cauchy jumps," *2009 IEEE Congress on Evolutionary Computation*, pp. 3285-3291, Trondheim, 2009.

[91] A. M. Duran-Rosal, P. A. Gutierrez, A. Carmona-Poyato, and C. Heryas-Martinez, "A hybrid dynamic exploitation bare-bones particle swarm optimisation algorithm for time series segmentation," *Neurocomputing*, vol. 353, pp. 45-55, Aug. 11, 2009.

[92] H. Wang, S. Rahnamayan, H. Sun, and M. G. H. Omran, "Gaussian bare-bones differential evolution," *IEEE Transactions on Cybernetics*, vol. 43, no. 2, Apr. 2013.

[93] X. Zhou, Z. Wu, H. Wang, and S. Rahnamayan, "Gaussian bare-bones artificial bee colony algorithm," *Soft. Comp.*, vol. 20, no. 3, pp. 907-924, 2016.

[94] L. Wang, W. Zhao, Y. Tian, and G. Pan, "A bare bones bacterial foraging optimization algorithm," *Cognitive Systems Research*, vol. 52, pp. 301-311,2018.

[95] H. Yu, W. Li, C. Chen, J. Liang, W. Gui H. Chen, and M. Wang, "Dynamic Gaussian bare-bones fruit fly optimizer with abandonment mechanism: method and analysis," *Engineering with Computers*, 2020.

[96] H. Peng, C. Deng, H. Wang, W. Wang, X. Zhou, and Z. Wu, "Gaussian bare-bones cuckoo search algorithm," in *Proceedings of the Genetic and Evolutionary Computation Conference 2018 (GECCO '18)*, Jennifer B. Sartor, Theo D'Hondt, and Wolfgang De Meuter (Eds.). ACM, New York, NY, USA, Article 4, pp. 1-2, 2018.

[97] H. Peng, and S. Peng, "Gaussian bare-bones firefly algorithm," *Int. J. Innovative Computing and Applications*, vol. 10, no. 1, pp. 35-42, 2019.

[98] M. Clerc, and J. Kennedy, "The particle swarm - explosion, stability, and convergence in a multidimensional complex space," in *IEEE Transactions on Evolutionary Computation*, vol.6, no. 1, pp. 58-73, Feb. 2002.

[99] F. van den Bergh, and A. P. Engelbrecht, "A study of particle swarm optimization particletrajectories," *Information Sciences*, vol. 176, no. 8, pp. 937-971, 2006.

[100] X.F. Wang, Y. Song, M. Irving, "Modern power system analysis," New York, NY, USA: Springer, 2008.

[101] R. D. Zimmerman, C. E. Murillo-Sanchez, and D. Gan, "*MATPOWER: A Matable Power System Simulation Package*," Accessed: May 4, 2019. [Online]. Available: http://www.pserc.cornell.edu/matpower/

[102] M. Ghasemi, M. M. Ghanbarian, S. Ghavidel, S. Rahmani, and E. M. Moghaddam, "Modified teaching learning algorithm and double differential evolution algorithm for optimal reactive power dispatch problem: A comparative study", *Inf. Sci.*, vol. 278, no. 10, pp. 231-249, Sep. 2014.

[103] T. Das, and R. Roy, "Optimal reactive power dispatch using JAYA algorithm," *Emerging Trends in Electronic Devices and Computational Techniques (EDCT)*, 2018.

[104] Y. Li, Y. Wang, and B. Li, "A hybrid artificial bee colony assisted differential evolution algorithm for optimal reactive power flow," *Electrical Power and Energy Systems.*, vol. 52, pp. 25-33, 2013.

[105] C. H. Liang, C. Y. Chung, K. P. Wong, and X. Z. Duan, "Comparison and improvement of evolutionary programming

techniques for power system optimal reactive power flow," *lEE Proc. Generation Transmission Distribution*, vol. 153, no. 2, pp. 228-236, 2006.

[106] C. H. Liang, C. Y. Chung, K. P. Wong, X. Z. Duan, and C. T. Tse, "Study of differential evolution for optimal reactive power flow", *IET Proc Gen Trans Distrib*, vol. 1, no. 2, pp. 253-260, 2007.

[107] C. Y. Chung, C. H. Liang, K. P. Wong, and X. Z. Duan, "Hybrid algorithm of differential evolution and evolutionary programming for optimal reactive power flow," *lET Generation Transmission and Distribution*, vol. 4, no. 1, pp. 84-93, 2010.

[108] S. Pandya, and R. Roy, "Particle swarm optimization based optimal reactive power dispatch," *IEEE Int. Conf. on Electrical Computer and Communication Technologies.*, pp. 1-5, 2015.

[109] P. Subbaraj, and P. N. Rajnarayanan, "Optimal reactive power dispatch with Cauchy and adaptive mutations," *International Conference on recent trends in Information Telecommunication and Computing IEEE Computer Society*, pp. 110-115, 2010.

[110] S. Mouassa, and T. Bouktir, "Artificial bee colony algorithm for discrete optimal reactive power dispatch," in *Proc. Int. Conf. Ind. Eng. Syst. Manage. (IESM)*, Oct. 2015, pp. 654-662, Oct. 2015.

[111] R. P. Singh, V. Mukherjee, and S. P. Ghoshal,"Optimal reactive power dispatch by particle swarm optimization with an aging leader and challengers," *Appl. Softw. Comput.*, vol. 29, pp. 298-309, 2015.

[112] A. Ghasemi, K. Valipour, and A. Tohidi, "Multi objective optimal reactive power dispatch using a new multi objective strategy," *Int. J. Electr. Power Energy Syst.,* vol. 57, pp. 318-334, 2014.

[113] S. Duman, Y. Sönmez, U. Güvenç, and N. Yörükeren, "Optimal reactive power dispatch using a gravitational search algorithm," *IET Gener. Transm. Distrib.*, vol. 6, no. 6, pp. 563-576, 2012.

[114] K. Mahadevan, and P.S. Kannan, "Comprehensive learning particle swarm optimization for reactive power dispatch", *Applied Soft Computin*g, vol. 10, pp. 641-652, 2010.

[115] Yang Li, Chunling Wang, Guoqing Li, and Chen Chen. "Optimal scheduling of integrated demand response-enabled integrated energy systems with uncertain renewable generations: A Stackelberg game approach," *Energy Conversion and Management*, 2021, 235: 113996.

[116] Yang Li, Meng Han, Zhen Yang, Guoqing Li. "Coordinating flexible demand response and renewable uncertainties for scheduling of community integrated energy systems with an electric vehicle charging station: A bi-level approach," *IEEE Transactions on Sustainable Energy*, 2021, 12(4): 2321-2331.

[117] Yang Li, Kang Li, Zhen Yang, Yang Yu, Runnan Xu, and Miaosen Yang. "Stochastic optimal scheduling of demand response-enabled microgrids with renewable generations: An analytical-heuristic approach," *Journal of Cleaner Production*. 2022, 330: 129840.
23